\documentclass[a4paper,aps,pre,showpacs,twocolumn,superscriptaddress,groupedaddress,longbibliography]{revtex4-1}

\usepackage{amsmath}
\usepackage{amssymb}
\usepackage{graphicx}
\usepackage{xcolor}
\usepackage{natbib}
\usepackage{siunitx}
\usepackage{appendix}

\usepackage{ulem}
\newcommand{\beq}{\begin{equation}}
\newcommand{\eeq}{\end{equation}}

\newcommand{\G}{\Gamma}
\newcommand{\Hinf}{H_\infty}
\newcommand{\hinf}{h_\infty}
\newcommand{\twait}{t_{\text{wait}}}
\newcommand{\thold}{t_{\text{hold}}}
\newcommand{\Wadh}{W_{\mathrm{sep}}}
\newcommand{\Hmin}{H_{\text{min}}}

\begin{document}

\title{Elasto-capillary adhesion: The effect of deformability on adhesion strength and detachment}
\author{Matthew Butler, Finn Box, Thomas Robert and Dominic Vella}
\affiliation{Mathematical Institute, University of Oxford, Woodstock Rd, Oxford OX2 6GG, United Kingdom}
\date{\today}

\begin{abstract}
We study the interaction between capillary forces and deformation in the context of a deformable capillary adhesive: a clamped, tense membrane is adhered to a rigid substrate by the surface tension of a liquid droplet. We find that the equilibrium adhesive force for this elastocapillary adhesive is significantly enhanced in comparison to the capillary adhesion between rigid plates. In particular, the equilibrium adhesion force is orders of magnitude greater when the membrane is sufficiently deformed to contact the substrate. From a dynamic perspective, however, the formation of a fluid-filled dimple slows this approach to contact and means that stable attachment is only achieved if adhesion is maintained for a minimum time. The inclusion of a variable membrane tension (as a means of modifying the deformability) gives additional control over the system, allowing new detachment strategies to be explored.
\end{abstract}

\pacs{47.55.nk; 46.70.Hg; 47.15.gm}

\maketitle

\section{Introduction}

Many insect species are capable of extraordinary feats of adhesion. They are able to climb smooth vertical surfaces, and have been observed supporting loads exceeding 100 times their body weight whilst upside-down and remaining adhered  to glass \cite{Dirks2011}. This adhesion occurs reliably on surfaces with various  surface chemistries and across a wide range of scales \cite{Labonte2015}, and the insects are able to adhere and detach continuously during locomotion. The current understanding is that this adhesion is mediated by  a secreted fluid  that attaches their feet to the substrate via capillary forces \cite{Dirks2014}.

Many technological adhesives have been motivated by such biological adhesives: for example, a switchable rigid capillary adhesive has been proposed taking insects as its inspiration \cite{Vogel2010}. Other examples include: a soft pressure-controlled pad that can deform to grasp complex surfaces \cite{Song2017} and a robotic gripper for use in capturing space-debris \cite{Jiang2017} (both of which make use of gecko-like microfibrils \cite{Autumn2002}), as well as an octopus-inspired patch that can be used to adhere underwater \cite{Baik2017}.

The simplest models of insects' `wet' adhesion are based on the confinement of a droplet between two rigid surfaces --- squashing the droplet in a narrow gap allows for the calculation of the adhesion force provided by capillary pressure for a Hele-Shaw cell \cite{Dirks2014,Reyssat2015} or for more complex geometries \cite{Butt2009}. Quantifying this underlying mechanism gives plausible values of the observed adhesive strength. However, detachment requires the droplet to be de-stabilized \cite{Lowry1995,Slater2014} and, if performed dynamically, can lead to large resistive viscous forces \cite{Bikerman1968}. 

While models of capillary adhesion between rigid surfaces are a useful starting point, it has been observed that the footpads of some insects are soft and deformable \cite{Gorb2000}. In this paper, we will consider how this deformability modifies the classic results attained for rigid surfaces. In particular, the coupling between deformability and adhesion may allow the surfaces to come into closer contact, increasing the maximum adhesive force. Moreover, if this deformability can be controlled, it may also allow for new mechanisms of detachment.

In recent years there has been significant progress on a variety of problems related to the interaction between deformable surfaces and surface tension, with so-called `elasto-capillary' systems exhibiting many interesting and counter-intuitive phenomena \cite{Roman2010,Style2017,Bico2018}. One common theme in these settings is the occurrence of hysteresis and rapid transitions between markedly different states, for example the zipping and unzipping of fibres by a droplet as the fibre tension or separation is varied \cite{Duprat2015}. Furthermore, theory and experiments suggest that fluid droplets are capable of significant deformation of surfaces: beams clamped at one end can be bent into contact \cite{Taroni2012,Kwon2008} and two soft elastic half-spaces can be pulled together \cite{Wexler2014} by the forces of a single fluid droplet.

Inspired by the adhesive capabilities of insects, we study a model elasto-capillary system: a tense membrane is adhered to a rigid, planar substrate by the action of a fluid droplet. A mathematical model of this system is outlined in section \ref{sec:model}, where we use a local force-balance to determine the governing equations, and highlight the key parameters of the system. In section \ref{sec:eql} we study the  equilibria of the system, focussing on the adhesion force and taking care to include solutions in which the membrane is deformed sufficiently to contact the substrate. Section \ref{sec:expts} presents experimental data that supports the equilibrium picture, but also hints at the importance of dynamics, which we then study in section \ref{sec:dynamics} using lubrication theory. We consider adhesive detachment in section \ref{sec:detach}, focussing on finding a strategy that minimizes the work required to release the membrane from the substrate. Finally, in section \ref{sec:concl}, we summarize our results and consider directions for future work.

\section{Model} \label{sec:model}

\begin{figure}[tbp]
	\centering
	\includegraphics[width=0.9\linewidth]{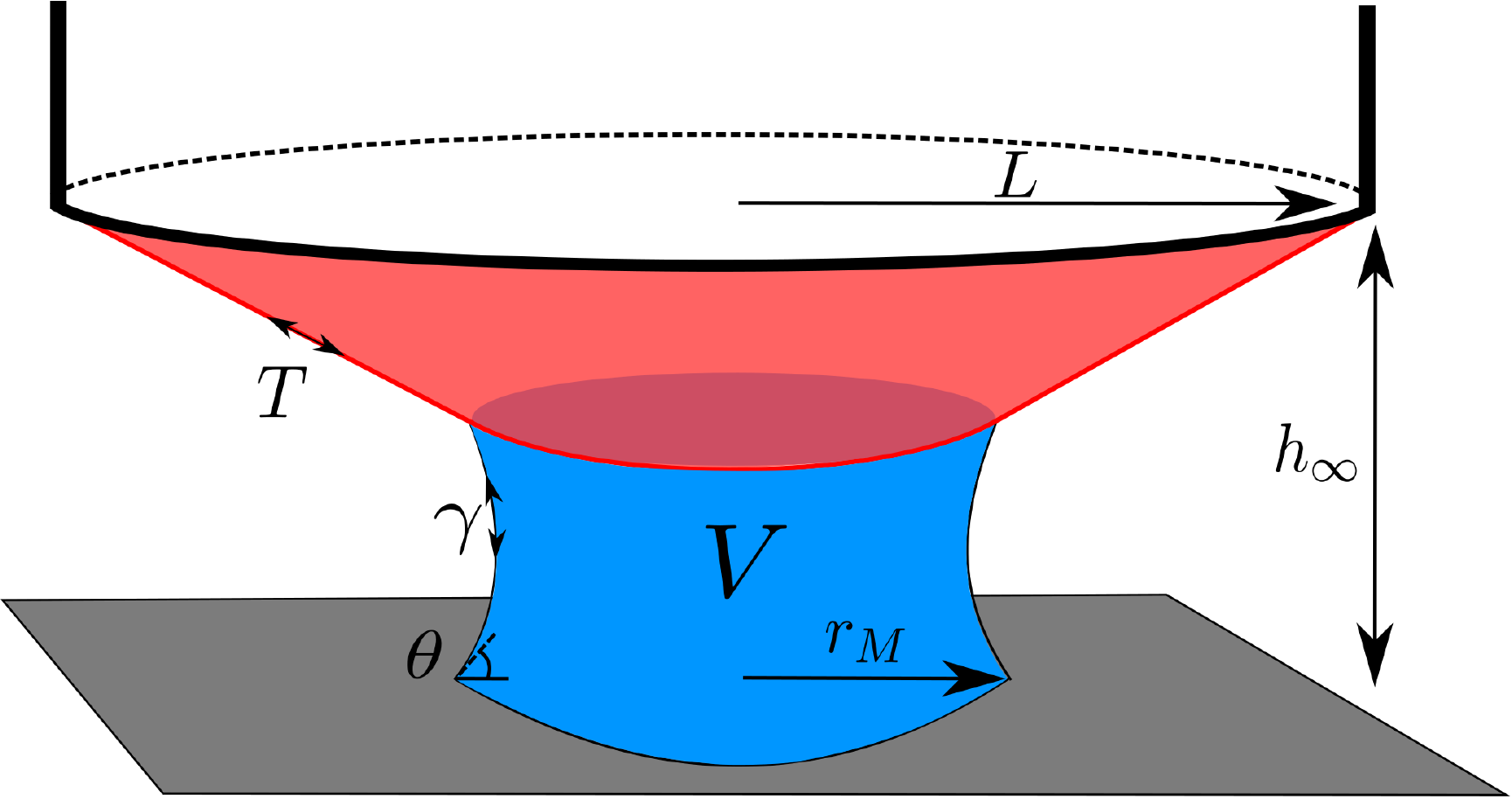}
	\caption{A liquid droplet (blue) of volume $V$ bridges the gap between a rigid plate (grey) and a deformable membrane (red). The membrane (of thickness $\tau$) is clamped around a circle of radius $L$ with an applied tension $T$ and at a height $\hinf$ above the plate. The free surface of the droplet has a surface tension $\gamma$ and makes a contact angle $\theta$ with each surface.}
	\label{fig:Setup}
\end{figure}

To investigate the possible role of elasticity in capillary adhesion, we consider perhaps the simplest deformable surface possible: a circular membrane of thickness $\tau$ and Young's modulus $E$. We shall consider only small axisymmetric deformations of the membrane, which is subject to a constant imposed tension and clamped at the radial position $r=L$. Adhesion to a flat and rigid target surface (which lies a distance $\hinf$ below the clamped edge of the membrane) is achieved by introducing a liquid droplet between the two surfaces (as shown in fig.~\ref{fig:Setup}). We anticipate that the two surfaces will generally be `close to contact' and hence that the aspect ratio is small, $\hinf/L\ll1$, though for clarity our figures will exaggerate the vertical scale; furthermore we neglect the effect of gravity on the droplet and membrane.

The surface tension of the droplet deforms the membrane because of the forces it exerts on the membrane. This deformation in turn modifies the surface tension force leading to novel feedbacks in this system; our aim is to understand how the interaction between capillarity and deformability affect the adhesive properties of the system, particularly in comparison to capillary adhesion of rigid surfaces. 

For simplicity, we shall take the tension $T$ in the membrane  to be uniform and treat $T$ as a control parameter: we neglect any modification of the uniform tension due to additional stretching of the membrane during vertical deformation. To understand when this approximation is valid, we note that simple geometry gives the deformation-induced strain  $\varepsilon \lesssim (\hinf/L)^2$. The thickness-integrated  stress in the membrane is then ${\sigma \sim T + E \tau \hinf^2/L^2}$: in scaling terms, the effect of deformation-induced stretching is negligible provided that ${T \gg E \tau(\hinf/L)^2}$.  (Note that this is a different condition from that for the pre-existing tension in a membrane to dominate that induced by a spherical cap droplet, $T\gg\gamma^{2/3}(E\tau)^{1/3}$ \cite{Davidovitch2018}.) Crucially, it is possible for the applied tension to dominate the geometry-induced tension even while the membrane remains Hookean; this requires that the tension-induced strain remains small, $T/(E\tau)\ll1$.  We shall also neglect the bending stiffness of the membrane, which requires that $T \gg E \tau^3/L^2$.

A droplet with interfacial tension $\gamma$ that is confined between two surfaces applies a capillary force to each surface in two distinct ways: (i) a jump in the normal stress (i.e.~a pressure difference between the inside and outside of the droplet) proportional to the meniscus curvature results in a force acting over the liquid-solid contact area, and (ii) a tension force as the meniscus pulls on the contact line where the membrane, liquid and vapour meet. When a wetting droplet is confined to a very narrow gap (and provided that the contact angle is not too close to $\pi/2$), the former dominates  the latter because the area-scaling of the pressure force `beats' the length-scaling of the line force. Additionally, at this contact line we expect the surface tension to contribute to a discontinuity in the membrane tension, $[T]^+_- \propto \gamma$, but we neglect this because of the high tensions considered, $\gamma \ll T$. We shall make these assumptions henceforth.

To quantify the adhesive force, we need to determine the interfacial curvature. Consistent with our assumption of a thin gap, we assume that the curvature is dominated by the component between the surfaces, and ignore the azimuthal contribution \cite{Reyssat2015}. Further, we approximate the meniscus cross-section as a circular arc \cite{Fisher1926} of radius $h_M/(2 \cos \theta)$, where $h_M$ is the height of the membrane above the substrate where the meniscus meets the membrane (or `meniscus height') and $\theta$ is the contact angle of the liquid-solid-vapour system, taken for simplicity to be the same on both surfaces. (We neglect any variations in the contact angle from the value given by the Young-Dupr\'{e} law \cite{deGennes2004}, because the applied tension $T\gg\gamma$ \cite{Davidovitch2018}. We also ignore the effect of the membrane slope at the contact line on the contact angle as the slopes are taken to be small, $h_r \ll 1$.) The pressure at the meniscus (which throughout this paper is measured relative to the ambient pressure) is then given by 
\begin{equation}
p_{\text{meniscus}} = - \frac{2 \gamma \cos \theta}{h_M}. 
\end{equation}

\subsection{The rigid case}

In what follows, we shall seek to understand the role of the membrane's deformability in modifying the adhesion force; it is therefore helpful to have the perfectly rigid case as a point of comparison, not least since we expect to recover this limit as the tension ${T\to\infty}$. In the rigid case the gap width is uniform, so that the meniscus height $h_M=\hinf$ and  ${p = - 2\gamma\cos\theta/\hinf}$. The adhesive force provided by a droplet of known volume $V$ is due to the pressure $2\gamma\cos\theta/\hinf$ acting over the area $A = V/\hinf$ (ignoring a small correction due to the meniscus shape) and thus the force is simply 
\beq 
f_{\text{rigid}}=2\gamma\cos\theta \frac{V}{\hinf^2}.
\label{eq:FTinfty}
\eeq 
In this rigid case, and for fixed drop properties (i.e.~fixed volume $V$ and surface tension $\gamma$) the adhesive force is solely controlled by the gap separation, $\hinf$.

\subsection{Deformation}

We assume that the droplet is positioned at the centre of the membrane and that the system is axisymmetric. (We expect that an off-centre droplet will move to the centre because of gradients in capillary pressure, in a similar manner to droplets in a rigid tapered channel, see \cite{Reyssat2014} for example.) The axisymmetric membrane position may then be written $z=h(r,t)$ with $r$ the radial coordinate and $t$ time. 

A local force-balance on the membrane requires that the membrane shape $h(r,t)$ must be determined as the solution of the Young--Laplace equation
\begin{equation}
\label{eq:Poisson}
T \nabla^2_r h = -p
\end{equation} where $\nabla_x^2$ denotes the axisymmetric Laplacian operator $\nabla_x^2 f = \frac{1}{x} \frac{\partial}{\partial x} (x \frac{\partial f}{\partial x})$ for any function $f(x,t)$, and $p(r,t)$ is the pressure field within the droplet, which is uniform in static scenarios but may vary spatially in the dynamic scenarios we consider in \S\ref{sec:dynamics}. Here we have assumed that the membrane slope remains small throughout (consistent with the small aspect ratio) and, further, neglect the membrane's inertia.

Finally, we impose a fixed droplet volume, $V$, \ through
\begin{equation}
V = 2 \pi \int_0^{r_M} r h ~\mathrm{d}r
\end{equation} 
where $r_M$ is the radial position of the meniscus. Note that we have again assumed the meniscus shape has negligible impact on the volume because of the small aspect ratio of the droplet.

\subsection{Non-dimensionalization}

We render the problem dimensionless using the extrinsic radial lengthscale $L$ and by rescaling the drop volume to unity,  introducing the vertical length scale $V/L^2$ in the process. Pressures, $p$, are rescaled by the typical Laplace pressure $\gamma L^2 \cos \theta /V$, and forces, $f$, are non-dimensionalized by $\gamma L^4 \cos\theta /V$. We therefore define dimensionless variables
\beq
\begin{aligned} \label{eq:NonDim}
&R=r/L,\quad H(R)=h(r) \times L^2/V, \\
P &= \frac{p}{\gamma L^2 \cos \theta /V}, \quad F = \frac{f}{\gamma L^4 \cos\theta /V}.
\end{aligned}
\eeq 
With this non-dimensionalization, the static membrane shape is controlled by two  dimensionless parameters
\begin{equation}
	\G = \frac{\gamma L^6 \cos \theta}{T V^2}, \qquad \Hinf = \frac{L^2 \hinf}{V}.
	\label{eq:DimlessVars}
\end{equation} 

Physically, the parameter $\Hinf$  represents the rescaled gap width and influences the system via the clamped boundary condition $H(1)=\Hinf$; varying $\Hinf$ corresponds to changing the separation of the membrane from the target surface. The  parameter  $\G$ represents the competition between a typical capillary force, $pr^2 \sim \gamma L^2/h \sim \gamma L^4/V$, pulling down on the membrane and a restoring tension force, $Tr ~\mathrm{d}h/\mathrm{d}r \sim TV/L^2$. We impose a tension $T\gg\gamma$; nevertheless $\G$ may remain an $O(1)$ quantity because of the amplifying effect of the ratio $L^6/V^2$ in \eqref{eq:DimlessVars}.

The parameter $\G$ can be considered a measure of the extent to which capillarity is able to deform the membrane: for $\G\ll1$, the membrane is relatively rigid and little deformation occurs while for $\G\gg1$ the membrane is highly deformed by capillarity. We therefore refer to $\G$ as the `deformability' of the membrane.

We note that in the relatively undeformable case, ${\G\ll1}$, we expect to recover the rigid result \eqref{eq:FTinfty}, which we write in dimensionless terms as
\beq \label{eq:RigidForce}
F_{\text{rigid}}=\frac{f_{\text{rigid}}}{\gamma L^4 \cos\theta /V}=\frac{2}{\Hinf^2}.
\eeq
To understand how the adhesion force deviates from this result as the deformability $\G$ increases, we turn to study the equilibrium problem. A key aim is to understand how the adhesive force $F(\G,\Hinf)$ behaves.

\section{Equilibrium Solutions} \label{sec:eql}

In equilibrium there is no fluid flow and so the internal droplet pressure must be uniform, and equal to the value at the meniscus, i.e.~${P=-2/H(R_M)}$ where $R_M$ denotes the radial position of the meniscus. The problem of determining the equilibrium membrane shape, $H(R;\G,\Hinf)$, therefore reduces to solving Poisson's equation \eqref{eq:Poisson} with forcing pressure
\beq
P(R)=\begin{cases}
-\frac{2}{H(R_M)},\quad &0\leq R<R_M\\
0,\quad\quad &R_M\leq R\leq 1.
\end{cases}
\eeq
The relevant boundary conditions are due to the imposed clamping (at $R=1$) and symmetry/regularity at the origin, i.e.~$H(1)=\Hinf$ and $H'(0)=0$. At the meniscus, the membrane height and slope are continuous. (In general, a horizontal force balance at the contact line between membrane, liquid and vapour shows that the membrane slope may have a discontinuity proportional to $\gamma \sin \theta$; this can be neglected provided that the droplet aspect ratio $r_M/h_M \gg \tan \theta$.) 

\subsection{Problem statement}

In dimensionless terms, the equilibria of the system satisfy
\begin{equation} \label{eq:EqlEqn}
	\nabla^2_{R} H = \begin{cases}
				2\G / H_M,  \qquad \qquad &0<R<R_M,\\
				 0, \qquad &R_M<R<1.
				\end{cases}
\end{equation}
The solution of \eqref{eq:EqlEqn} is to be found subject to the boundary conditions
\begin{align}
	\left[ H \right]_{-}^{+} &= \left[\frac{\mathrm{d}H}{\mathrm{d}R}\right]_-^+ = 0,  &R=R_M, \\	
	&\frac{\mathrm{d}H}{\mathrm{d}R}=0,  &R=0, \label{eq:NoContactBC} \\
	&H = \Hinf,	 &R=1.
\end{align}
Note that the radial position of the meniscus, $R_M$, and its height, $H_M$, are not known \emph{a priori} and must be determined as part of the solution. We therefore require two additional relations. The first of these is simply that $H(R_M)=H_M$. The second, and final, condition is the imposed volume constraint, namely
\begin{equation}
	1 = 2 \pi \int_0^{R_M} R H ~\mathrm{d}R.
	\label{eq:VolND}
\end{equation}

For given values of $\Hinf$ and $\G$, the system \eqref{eq:EqlEqn}--\eqref{eq:VolND} may be solved analytically to give a single transcendental equation for $R_M$: 
\begin{equation}
\begin{aligned}
	(\pi H_{\infty} R_M^2 -1)^2 +& (1 - 4\log{R_M})(\pi H_{\infty} R_M^2 -1) \\
	-&\frac{\pi^2}{4}\G\,(1 - 4\log{R_M})^2  R_M^6  = 0.
\end{aligned}
\label{eq:RmTrans}
\end{equation} This equation could be rearranged to give $\Hinf$ for given $R_M$, but we prefer to  solve \eqref{eq:RmTrans} numerically to find the (unknown) radius $R_M$ for given $\G$ and $\Hinf$, subject to the constraint that the liquid must remain within the domain, i.e.~${R_M<1}$. 

We find numerically that for some parameter values, the membrane touches the lower plane, i.e.~$H=0$ at some radial position $R$, which is generally $R=0$ (since the membrane is most deformable in the centre). When this happens, the nature of the solution changes (since  the membrane cannot penetrate the base, $H(R) \ge 0 $ for all $ 0\leq R\leq 1$); we therefore consider contacting solutions separately now.

\subsection{Contacting solutions}

When the membrane is in contact with the rigid surface in some region $R<C$, it is no longer solely subject to the capillary pressure but  also to an unknown reaction force provided by the base. In this contacting region, the shape of the membrane is therefore no longer governed by \eqref{eq:Poisson} but rather by the requirement that  the membrane conforms to the base, i.e.~$H=0$. At the boundary between contacting and non-contacting regions, a local force balance reveals that the membrane height and gradient should be continuous (assuming that contact does not give rise to additional adhesion or repulsion). We therefore require that at the edge of the solid--solid contact region (i.e.~$R=C$): $H=\mathrm{d}H/\mathrm{d}R=0$. Note that with this condition, it is not possible to have an equilibrium with an annular contact containing trapped fluid.

The contacting problem is therefore largely the same as the non-contacting problem, save that $H=0$ for $R<C$. The most significant change is that the wet region is now $C<R<R_M$ with $C$  an additional unknown to be found. The additional constraint required to find $C$ comes from  the two continuity conditions at the edge of the solid--solid contact region 
\begin{equation}
	H = \frac{\mathrm{d}H}{\mathrm{d}R}=0, 	\qquad \qquad   R=C,
\end{equation} which replace the symmetry boundary condition of \eqref{eq:NoContactBC}, $H'(0)=0$.

More concretely, the contact problem reduces to solving the following three non-linear simultaneous equations for $R_M, H_M$ and $C$
\begin{align}
H_M &= \frac{\G}{2H_M} \left[ R_M^2 - C^2 + 2C^2 \log{(C/R_M)} \right], \label{eq:ContactCondition} \\
	H_{\infty} &= H_M - \frac{\G}{H_M}(R_M^2-C^2) \log{R_M}, \label{eq:ContactGrad} \\
	1 &= \frac{\pi \G}{H_M} \left[ \frac{1}{4}(R_M^4 - C^4) + C^2R_M^2 \log{(C/R_M)} \right]. \label{eq:ContactVol}
\end{align}

\subsection{Adhesion Force and Multiple Solutions}

\begin{figure}[tbp]
	\centering
	\includegraphics[width=0.9\linewidth]{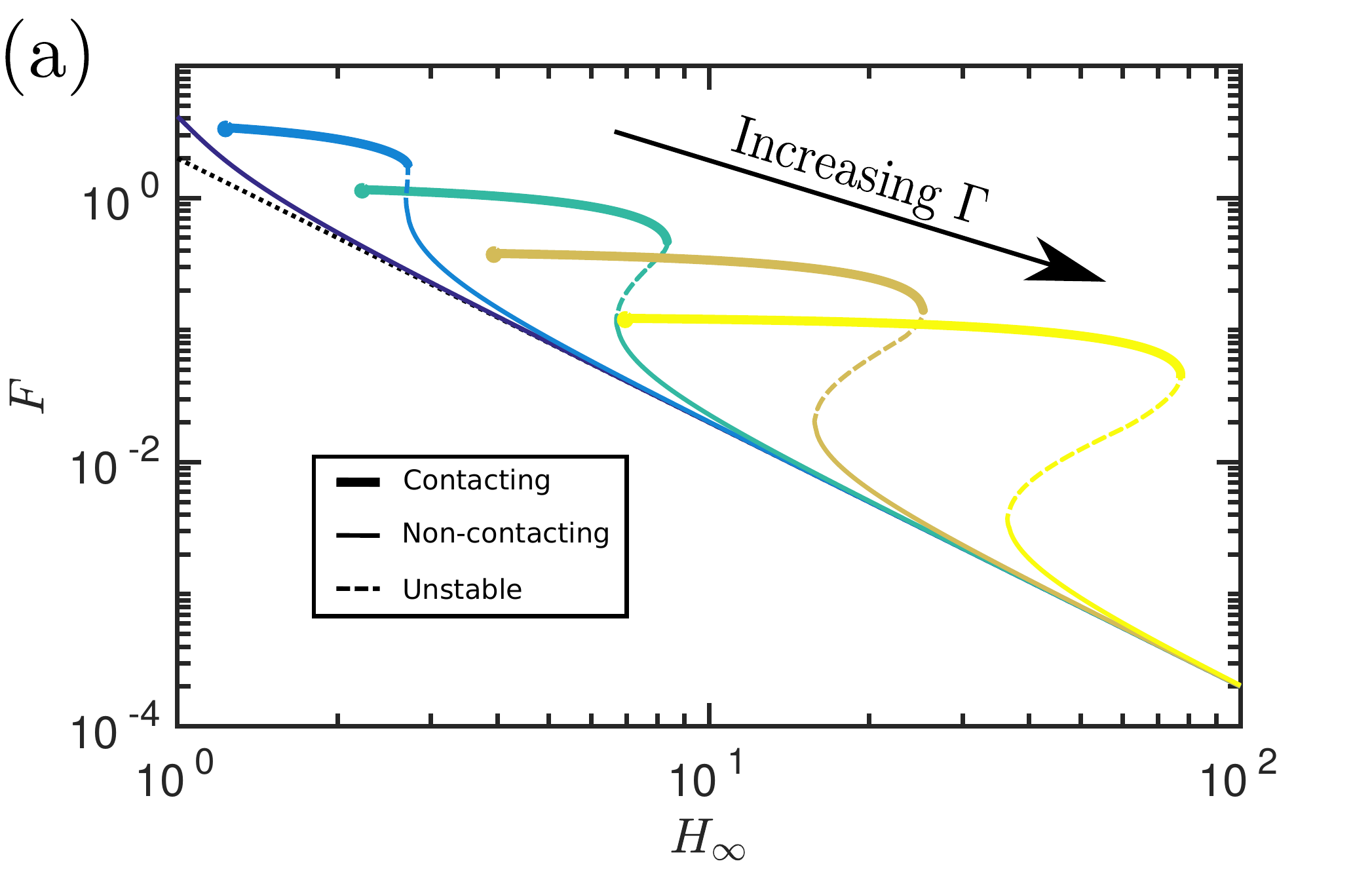}
	\includegraphics[width=0.9\linewidth]{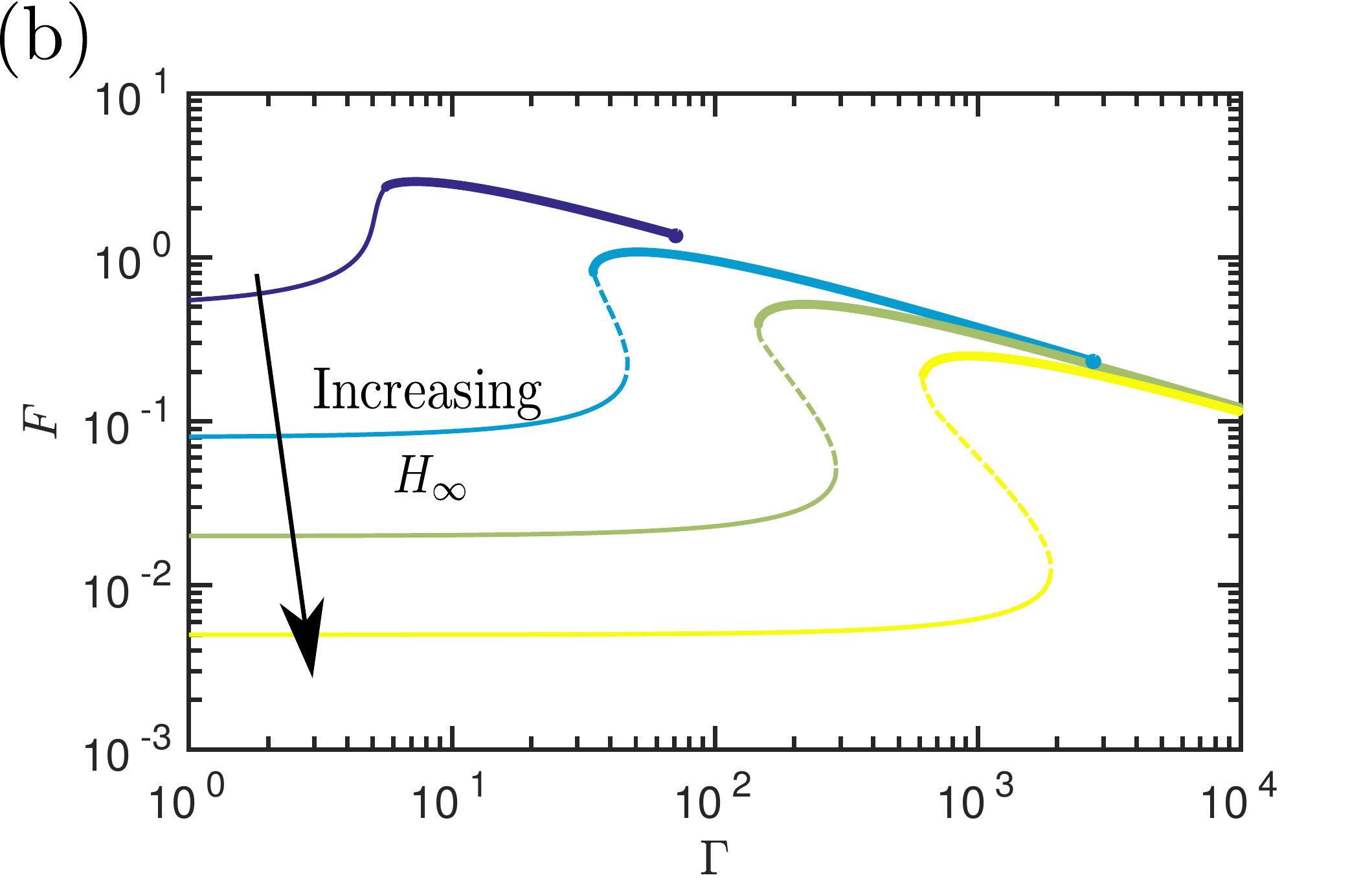}
	\caption{The dimensionless adhesive force $F(\G,\Hinf)$. Results are shown for: (a) fixed deformability ${\G=1,10,10^2,10^3,10^4}$ and varying edge height $\Hinf$ and (b) fixed edge height ${\Hinf=2,5,10,20}$ and varying membrane deformability $\G$. The computation stops when the system is flooded, i.e.~${R_M=1}$ (solid circles). In (a) the force is compared to a perfectly rigid membrane $\G=0$ (black dotted line) and in both panels different thickness curves are used to distinguish different states, as described in the legend of (a). We note that the adhesion force is significantly larger when the membrane is in contact with the base (thicker solid curves) but even out of contact (thinner solid curves) the adhesive force remains  larger than in the rigid case. }
	\label{fig:EqlForce}
\end{figure}

The adhesive force in  equilibrium, that is the force acting normal to the rigid surface (or equivalently the force that must be applied at the clamps to maintain the equilibrium) is readily determined to be:
\begin{equation}
	F = \begin{cases}
		2 \pi ~\frac{R_M^2}{H_M} \qquad &\text{non-contacting},\\
		\\
		2 \pi ~\frac{R_M^2-C^2}{H_M}	\qquad &\text{contacting}.
		\end{cases}
		\label{eq:AdhesForce}
\end{equation} 
Therefore, to calculate the adhesive force, all that is required is a solution of the transcendental equations for the equilibrium meniscus position $R_M$ and height $H_M$ (as well as the edge of the solid--solid contact region, $C$, if it exists), which may readily be found numerically. Substituting these values into \eqref{eq:AdhesForce}  we find the dimensionless adhesive force, $F$, as a function of the gap separation $\Hinf$ when $\G$ is fixed (fig.~\ref{fig:EqlForce}a); alternatively $F$ may be plotted as a function of the deformability $\G$ when $\Hinf$ is fixed (fig.~\ref{fig:EqlForce}b).

In fig.~\ref{fig:EqlForce}a, the edge height is varied at several (fixed) values of the deformability $\G$ and the resulting adhesion force is plotted alongside the force in the rigid case \eqref{eq:RigidForce}. At each value of the edge height the adhesion force is larger for a deformable membrane (regardless of the applied tension $\G$) than for  the  rigid case $\G=0$. The soft adhesive force can be as much as two orders of magnitude larger than the corresponding rigid adhesive force. Similarly the softer membranes are able to achieve a given adhesion force (for example to support a given load) at a larger gap separation. This suggest that the addition of deformability into a capillary adhesive may improve its adhesive capabilities significantly.

\begin{figure}[tbp]
	\centering
	\includegraphics[width=0.8\linewidth]{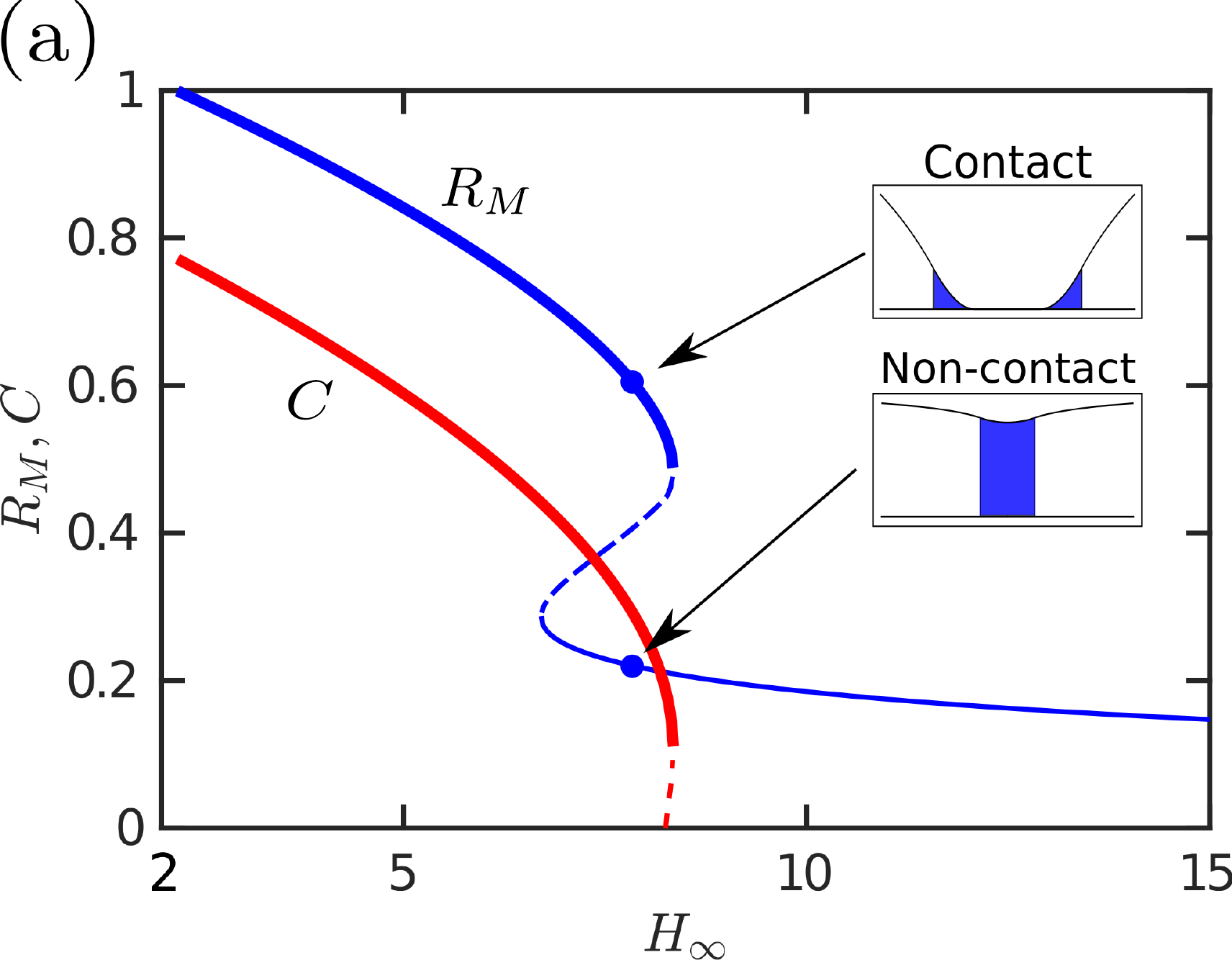}
	\includegraphics[width=0.8\linewidth]{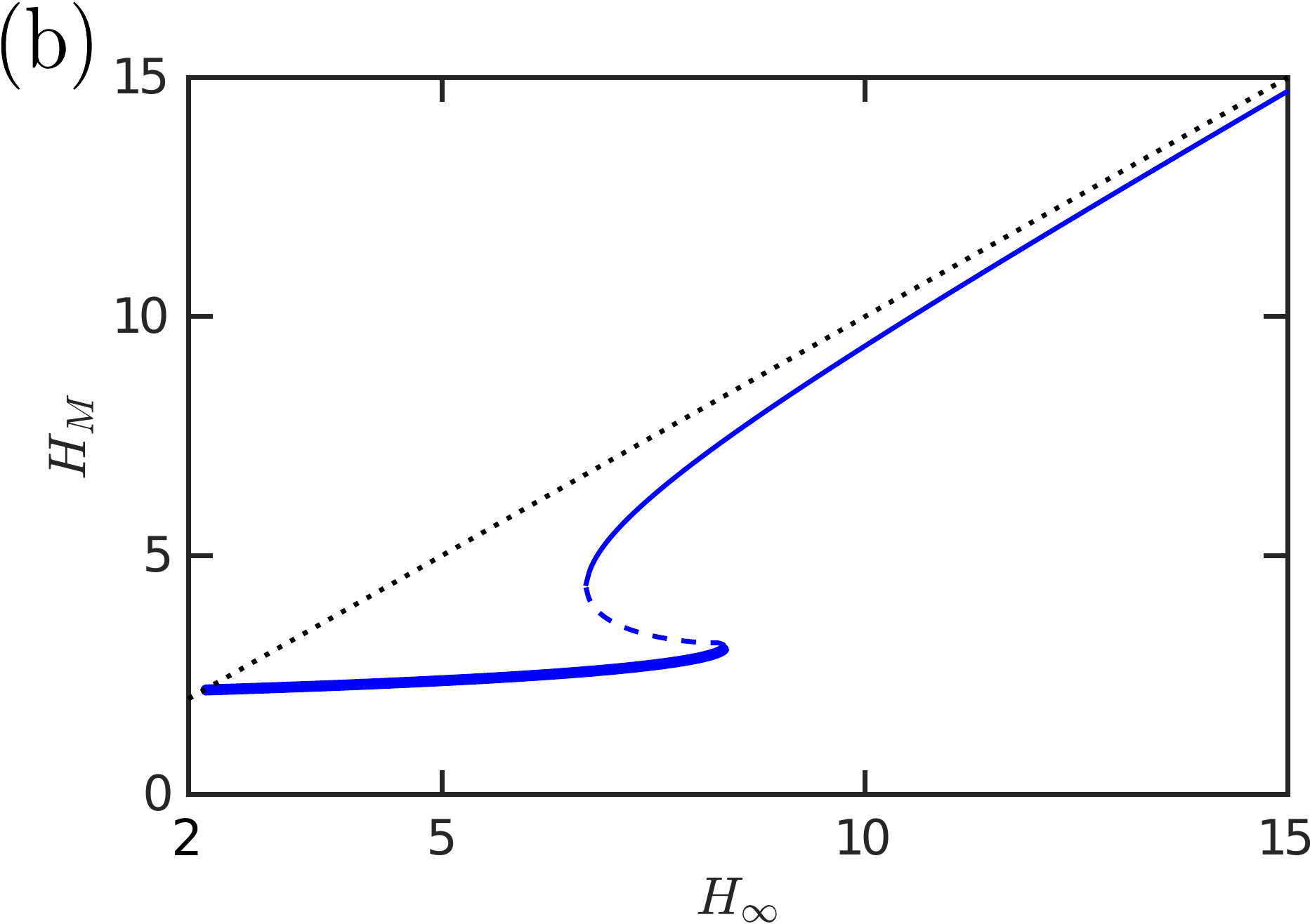}
	\caption{ (a) The equilibrium meniscus radius $R_M$ (blue) and contact point $C$ (red) vary with gap width $\Hinf$, when $\G$ is fixed (here $\G=100$). There is a non-contacting stable solution (thin solid curve), an unstable solution (dashed curve) and a contacting stable solution (thick solid curve). Each of these three states only exists over a specific range of values of $\Hinf$. Inset: Membrane profiles of the contacting and non-contacting stable solutions when $\Hinf = 8$ (blue circles).  
	(b) The variation of the meniscus height $H_M$ with $\Hinf$ (again shown for $\G=100$). The dotted line denotes $H_M=\Hinf$ for comparison.}
	\label{fig:Bifurcation}
\end{figure}

The behaviour of the solution, and in particular its adhesion force, is strongly characterized by whether contact occurs or not: when contact occurs, the droplet spreads further (fig.~\ref{fig:Bifurcation}a) and the meniscus height is significantly smaller (see fig.~\ref{fig:Bifurcation}b, as well as the two profiles that contrast the contacting and non-contacting states shown in the inset to fig.~\ref{fig:Bifurcation}a). Although the width of the wetted region, $R_M-C$, appears to be approximately constant (see fig.~\ref{fig:Bifurcation}a) the net result of the spreading in contact is an increase in the droplet footprint area. The two effects of an increased Laplace pressure and a larger area over which it acts lead to the dramatic increase in the adhesive force in the case of contact and motivate a more detailed study of \emph{when} contact occurs.

\begin{figure}[tbp]
	\centering
	\includegraphics[width=0.8\linewidth]{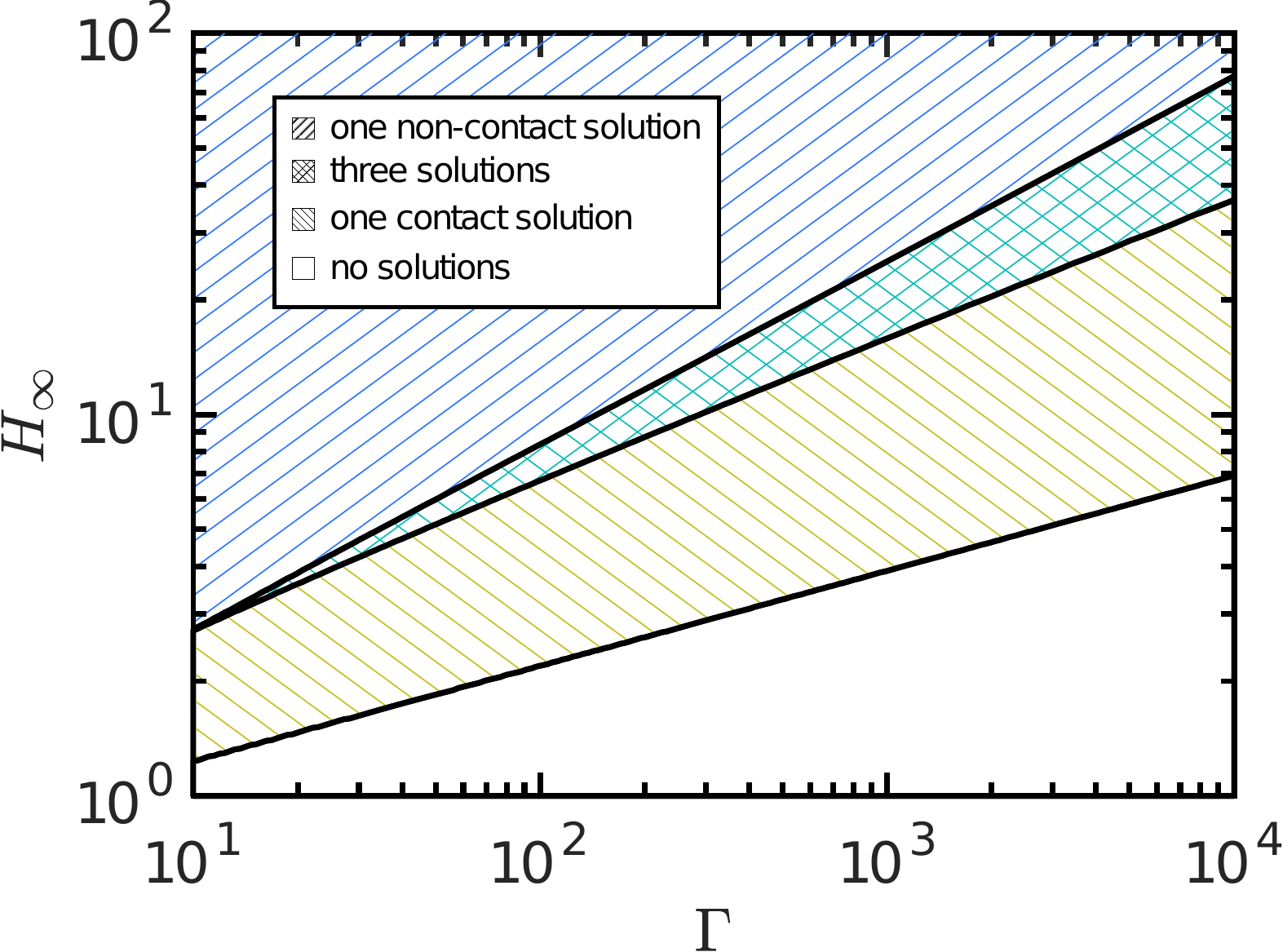}
	\caption{ The number and type of equilibria varies with the two parameters $\G$ and $\Hinf$. The hatching in each region denotes the number of contacting or non-contacting solutions, as described in the legend.}
	\label{fig:Multiple Solns}
\end{figure}

Exploring the $(\G,\Hinf)$ parameter-space for the number and type of equilibrium solutions we find three key regions: one in which there is a single non-contacting solution, another with a single contacting solution, and one region where three solutions are possible (of these three, one is contacting, one non-contacting and the third can be either, depending on the parameter choice). Fig.~\ref{fig:Multiple Solns} summarizes which regions of parameter space each behaviour is observed in. A stability analysis using Maddocks' theorem \cite{Maddocks1987} reveals that, in the case of 3 solutions, the intermediate solution is linearly unstable; the remaining two equilibria are linearly stable and consist of one contacting and one non-contacting solution. (Note that when $\Hinf$ becomes sufficiently small then the droplet floods the system and there is no longer any physically relevant solution.) 

In this system we have imposed a fixed separation $\Hinf$ and calculated the force $F$ that is generated; alternatively we could load the system with a given force $F$ and determine the corresponding $\Hinf$. In such a force-controlled scenario we expect all solutions to be unstable, with a small perturbation to the droplet or membrane resulting in either attachment with $\Hinf \to 0$ or detachment with $\Hinf \to \infty$ (see e.g.~\cite{Slater2014,Macner2014} for similar results in the rigid case).

Note that from fig.~\ref{fig:EqlForce}b it appears that at a given edge separation in the contacting regime, the force decreases as the strength of surface tension ($\gamma$) increases. However, because the force has been non-dimensionalized by the surface tension \eqref{eq:NonDim}, in fact the \emph{dimensional} force \emph{increases} as the surface tension increases, as expected.

The transition between non-contacting and contacting states (and vice versa) is sharp as the parameters are varied (figures \ref{fig:EqlForce}, \ref{fig:Bifurcation} \& \ref{fig:Multiple Solns}); this transition is a saddle-node bifurcation and introduces hysteresis into the system. The system can therefore be thought of as being `switchable': varying the parameters can `turn on' and `turn off' contact (and hence strong adhesion) as these transition points are passed. As we shall see shortly, this gives us alternative routes to de-adhere from the substrate: one can either increase $\Hinf$ fixing $\G$ (`yanking') or reduce $\G$ while maintaining $\Hinf$ (`peeling'). Before discussing this in more detail, however, we consider an experimental realization of the system discussed so far.

\section{Experiments} \label{sec:expts}

Our study of the equilibrium states of the system reveals two interesting features. Firstly, the system exhibits bi-stability --- for the same parameters (namely $\G$ and $\Hinf$) the system may be in one of two stable equilibrium states; secondly, the adhesion force of these two states may differ by more than an order of magnitude. To test whether these two states are physical, and to confirm the large difference in force between them, we developed an experimental version of this simple adhesive system.

\subsection{Set-up}

\begin{figure}[tbp]
	\centering
	\includegraphics[width=\linewidth]{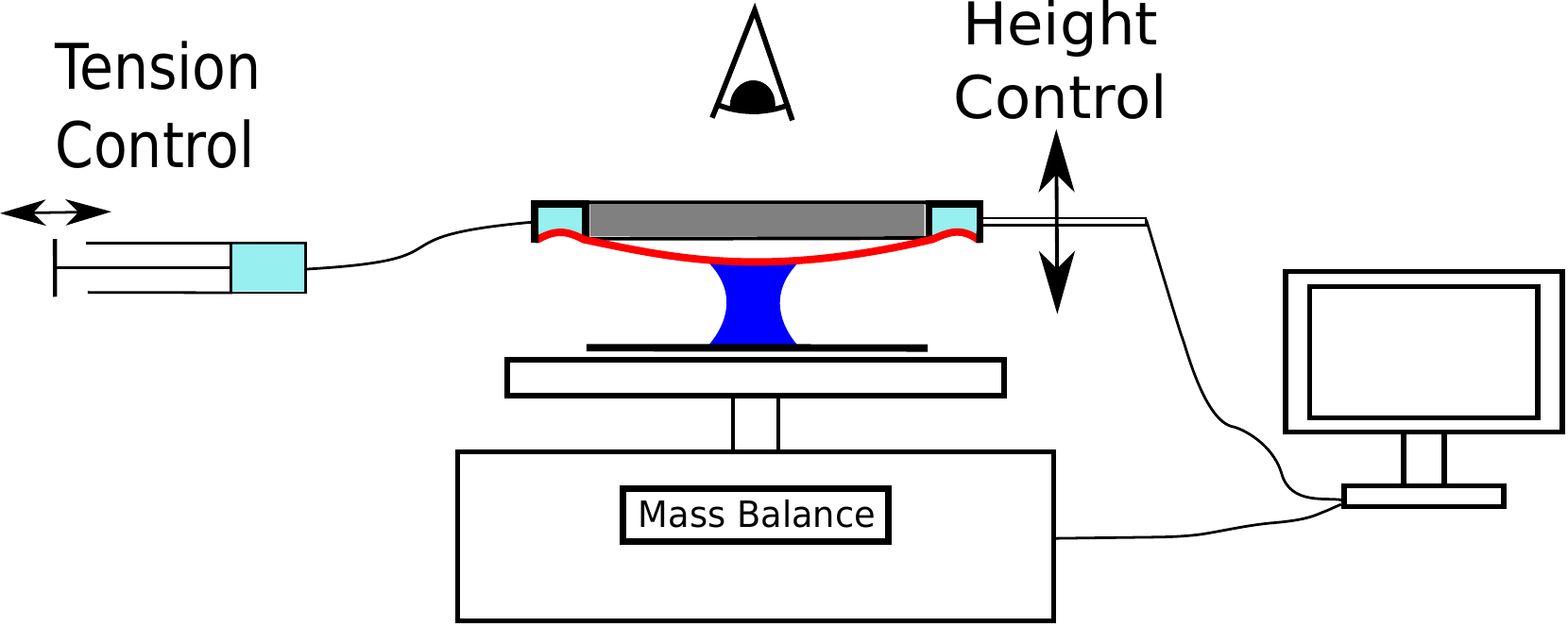}
	\caption{Schematic of the experimental set-up. A dyed droplet of mineral oil (blue) was confined between a PVS-coated glass plate and a clamped PVS sheet (red). This system rested on a mass balance that recorded the adhesive force. The height of the clamp (grey) was varied using a computer-controlled linear actuator, and the tension in the sheet could be increased by withdrawing air from an annular chamber (light blue) in the clamp, which sucked the outer edge of the sheet into the chamber, pulling the whole membrane taut. A camera imaged the experiment from above. } 
	\label{fig:ExptSetup}
\end{figure}

A thin (thickness $\tau\sim\SI{100}{\micro \meter}$), soft (Young's modulus $E=\SI{200}{\kilo \pascal}$) circular  sheet of Poly-vinyl siloxane (PVS) was fabricated by spin-coating. The sheet was clamped onto an annular chamber with radius $L = \SI{15}{\milli \meter}$;  the sheet tension was varied by withdrawing air from the chamber using a syringe, this created a pressure difference that sucked the outer edge of the membrane into the chamber, pulling the entire sheet taut across the clamp (see fig.~\ref{fig:ExptSetup}). The value of the tension was inferred by an indentation technique \cite{Vella2017}  prior to each experiment; when the tension was varied during the experiment, its value was also measured after the experiment. To ensure that the surface properties of the target surface were the same as that of the membrane (and, in particular, that the contact angles were the same), a rigid glass plate was coated with a layer of PVS with thickness $\sim100$\,$\mu$m to form the substrate. A dyed droplet of oil (Mineral Oil light, Sigma-Aldrich, UK; volume $1 \mathrm{~\mu L}\lesssim V\lesssim 10\mathrm{~\mu L}$) was confined between the clamped sheet and the PVS-covered glass, which itself rested on a mass balance accurate to 0.1mg (Pioneer PA64C Analytic Balance, Ohaus, Switzerland); this arrangement allowed the whole system to be weighed and the adhesive force determined to $1$\,$\mu$N precision. Mineral oil was chosen to reduce the effects of evaporation. The height of the clamped membrane was varied using a linear stage (M228.10S, Physik Instrumente, Germany) driven by a stepper motor (Mercury Step C663.11, Physik Instrumente, Germany), with a combined accuracy of $\pm2$\,$\mu$m. The force $f$ and plate separation $\hinf$ were both recorded digitally in MATLAB; typical measurements were in the range 0.1--50\,mN and 0.4--2\,mm, respectively.

A camera positioned above the experiment recorded the droplet's shape in plan view through the elastic sheet. The droplet radius was determined by least-squares fitting of a circular profile to the droplet's edge. The surface tension of the mineral oil was measured to be $ {\gamma = 32.1 \pm 0.2~\SI{}{\milli \newton \per \meter}}$ via the Wilhelmy plate method while the contact angle on PVS was measured to be $\theta=23.5 \pm \ang{2.5}$.

With these values, the tension-dominated regime of interest occurs when ${T \gg T_c \equiv E \tau \hinf^2/L^2}$; here ${T_c \le \mathcal{O}(0.1)\SI{}{\newton \per \meter}}$. Our experiments were conducted with the aspect ratio $\hinf/L \le~\mathcal{O}(0.1)$ and tensions in the range $1\mathrm{~N\,m^{-1}}\lesssim T\lesssim 10\mathrm{~N\,m^{-1}}$ so that $\gamma/T \lesssim 3\times 10^{-2}$. Therefore these experiments do indeed satisfy the various assumptions made to simplify the theoretical analysis (i.e.~$\hinf \ll L$, $\gamma \ll T$,  $T \gg E \tau(\hinf/L)^2$ and $T \gg E \tau^3/L^2$).

\subsection{Loading protocol}

Experiments were initially performed at fixed sheet tension, varying the gap width. A droplet of the mineral oil was placed on the PVS-covered glass. The clamped membrane was then lowered towards the droplet, with droplet contact detected by a sudden jump in the weight of the system (since the glass plate was partially lifted by the adhesive force of the liquid bridge as soon as it formed). 

Once we had detected that the droplet had bridged the membrane-substrate gap, there were three key stages to the experiment: (i) the clamp was lowered in steps of a few $\SI{}{\micro \meter}$, being left to settle on a timescale of $\SI{100}{\second}$  between steps, until the force began to evolve dynamically (when an inflection point was seen in the real-time measured force as a function of time), at which point (ii) the separation ($\hinf$) was kept constant for approximately 10 minutes to allow the system time to equilibrate, before (iii) retracting at a constant speed ($\SI{5}{\micro \meter \per \second}$) until the droplet bridge ruptured. The force was recorded throughout this process via changes in the  weight recorded by the mass balance, and the radius inferred from processing of images taken from above. The separation distance was determined from readings of the height of the clamp relative to the position at which dry surfaces contacted (measured prior to introduction of the droplet).

Further experiments were performed to understand the effect of changing tension. For these, we followed the procedure as above but once strong adhesion was achieved at the end of stage (ii), we increased the sheet tension over a period of $O(10\mathrm{~s})$,  instead of retracting the clamp. 

\subsection{Results}

\begin{figure}[tbp]
	\centering
	\includegraphics[width=0.8\linewidth]{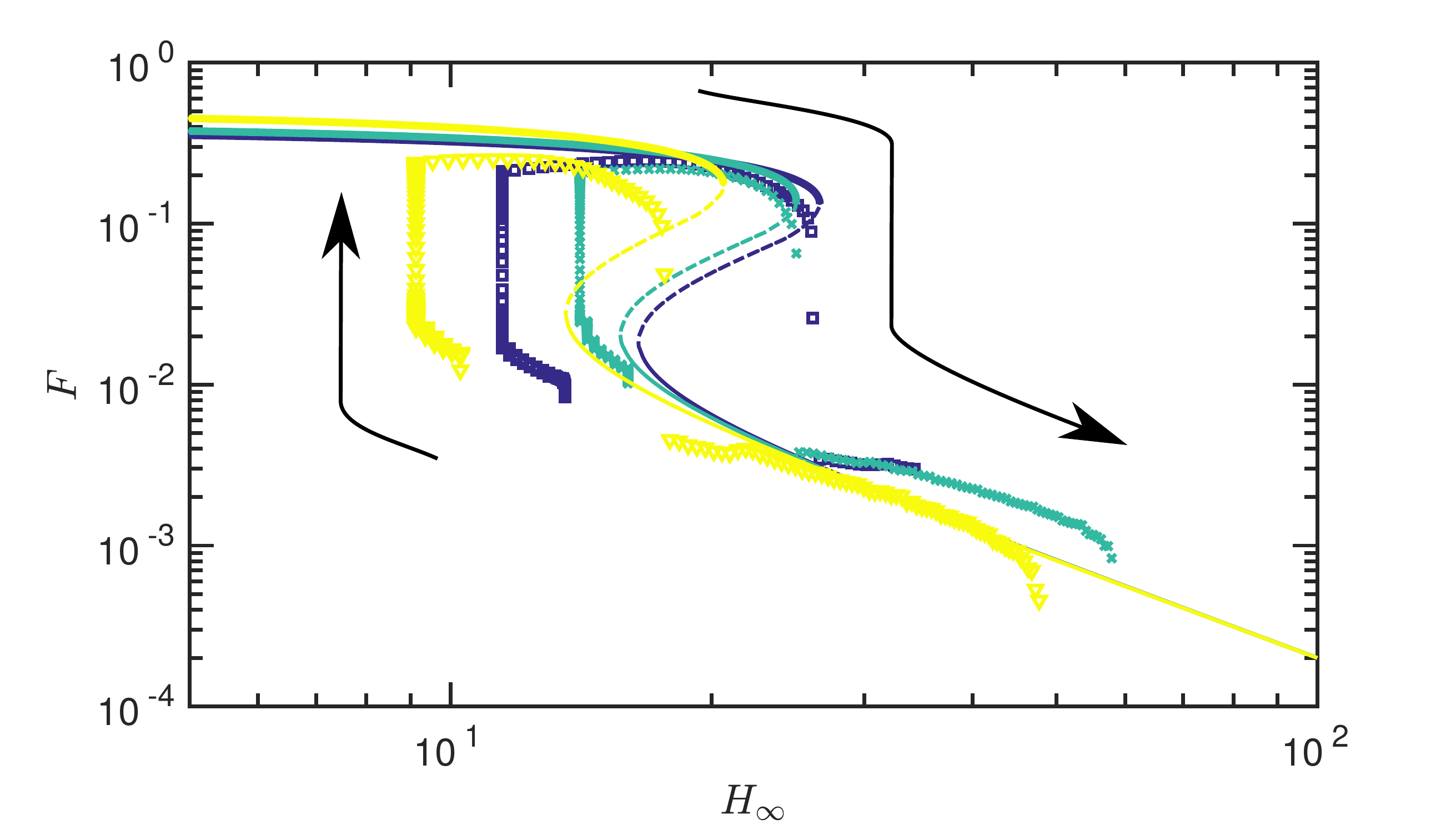}
	\includegraphics[width=0.8\linewidth]{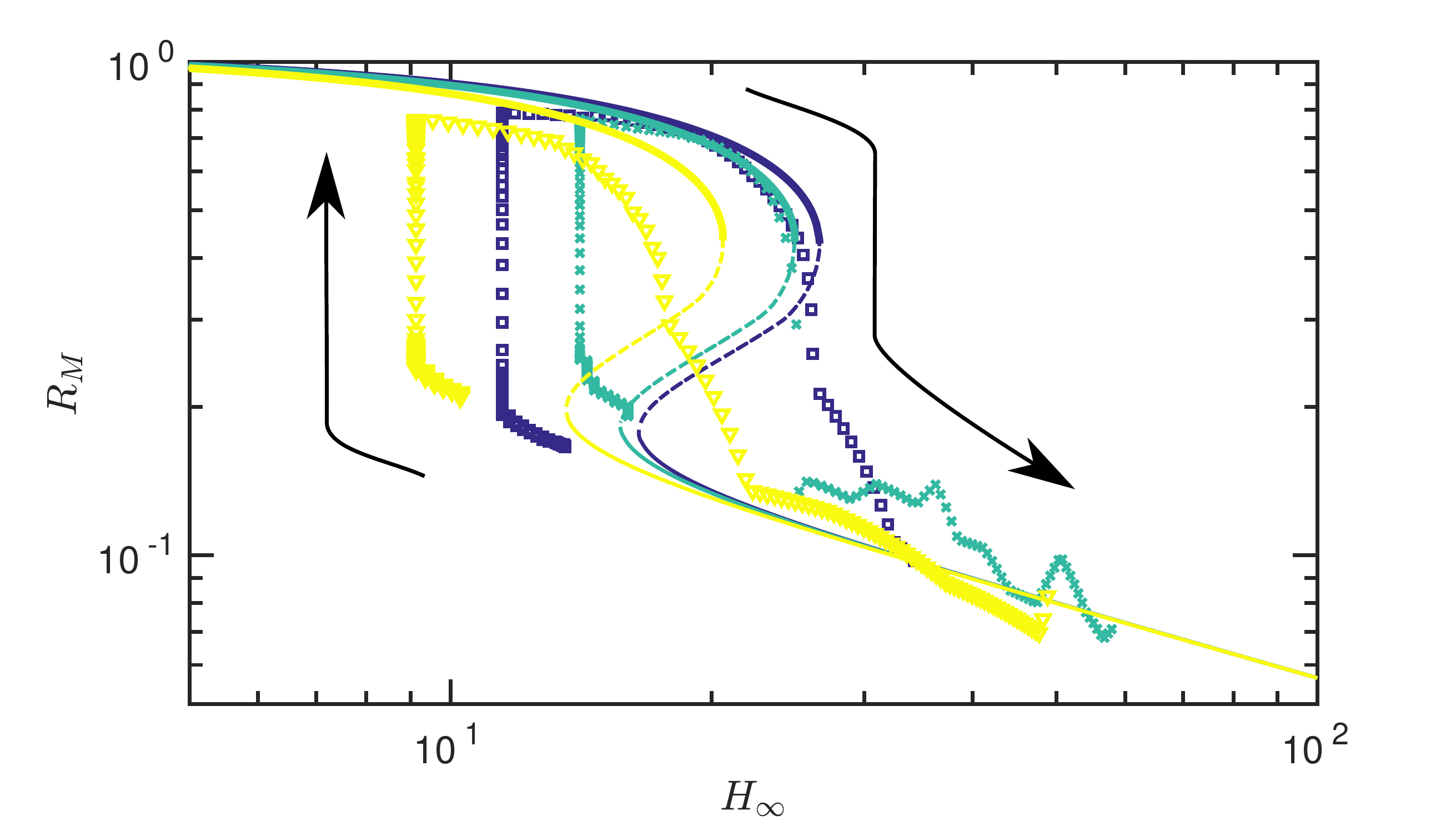}
	\caption{The measured dimensionless force and radius (symbols) compared with the theoretical predictions without any fitting parameter (curves) for three different experiments. The arrows illustrate the progression of the experiment: an initial lowering of the membrane, followed by a period at fixed height and then retraction at constant speed. Dark blue squares: $T = \SI{4.9}{\newton \per \meter}$, $V = \SI{7.8}{\micro \liter}$, $\G = 1100$. Blue-green crosses: $T = \SI{5.1}{\newton \per \meter}$, $V = \SI{8.2}{\micro \liter}$, $\G = 970$. Yellow triangles: $T = \SI{5.2}{\newton \per \meter}$, $V = \SI{9.9}{\micro \liter}$, $\G = 650$.}
	\label{fig:Experiment}
\end{figure}

Fixing the tension in the sheet and varying the gap width, we obtain force-displacement and radius-displacement curves, which are presented in fig.~\ref{fig:Experiment} with the corresponding (static) theoretical prediction with no fitting parameters.  We note that  the experiment exhibits the same phenomenological behaviour as predicted by our theory: there are two different stable states, each with markedly different adhesive force (and fluid extent) and, further, the transition between these two states is sharp as the gap width is varied. This transition occurs at different values of $\Hinf$ depending on the current state, i.e.~the system exhibits hysteresis. The critical parameter values at which the transition occurs, as well as the magnitude of the force and radius jump are predicted well by the equilibrium theory. This transition is noticeably much slower when the force is increasing towards contact than when the strong adhesion solution is lost (note that in fig.~\ref{fig:Experiment} data points are shown at constant intervals of 2.5~s, with many more points plotted during the motion into contact than the motion out of contact). We shall show in \S \ref{sec:dynamics} that the slowness of approach to contact is due to fluid being trapped beneath the membrane as contact is approached.

There are, however, some discrepancies between the theory and experiment which could be explained by factors such as: a misalignment of $\Hinf=0$ (calibrated before the droplet is added), dynamic effects, and additional forces not included in the model. As an example of an effect not included in the model, we note that at larger gap separations we start to move out of the small aspect ratio regime required by the theory (leading to the droplet necking and eventually rupturing); furthermore, the small drop radius in this case also means that the measurements of the radius become more unreliable as small errors in the fitting are more pronounced. 

\begin{figure}[tbp]
	\centering
	\includegraphics[width=\linewidth]{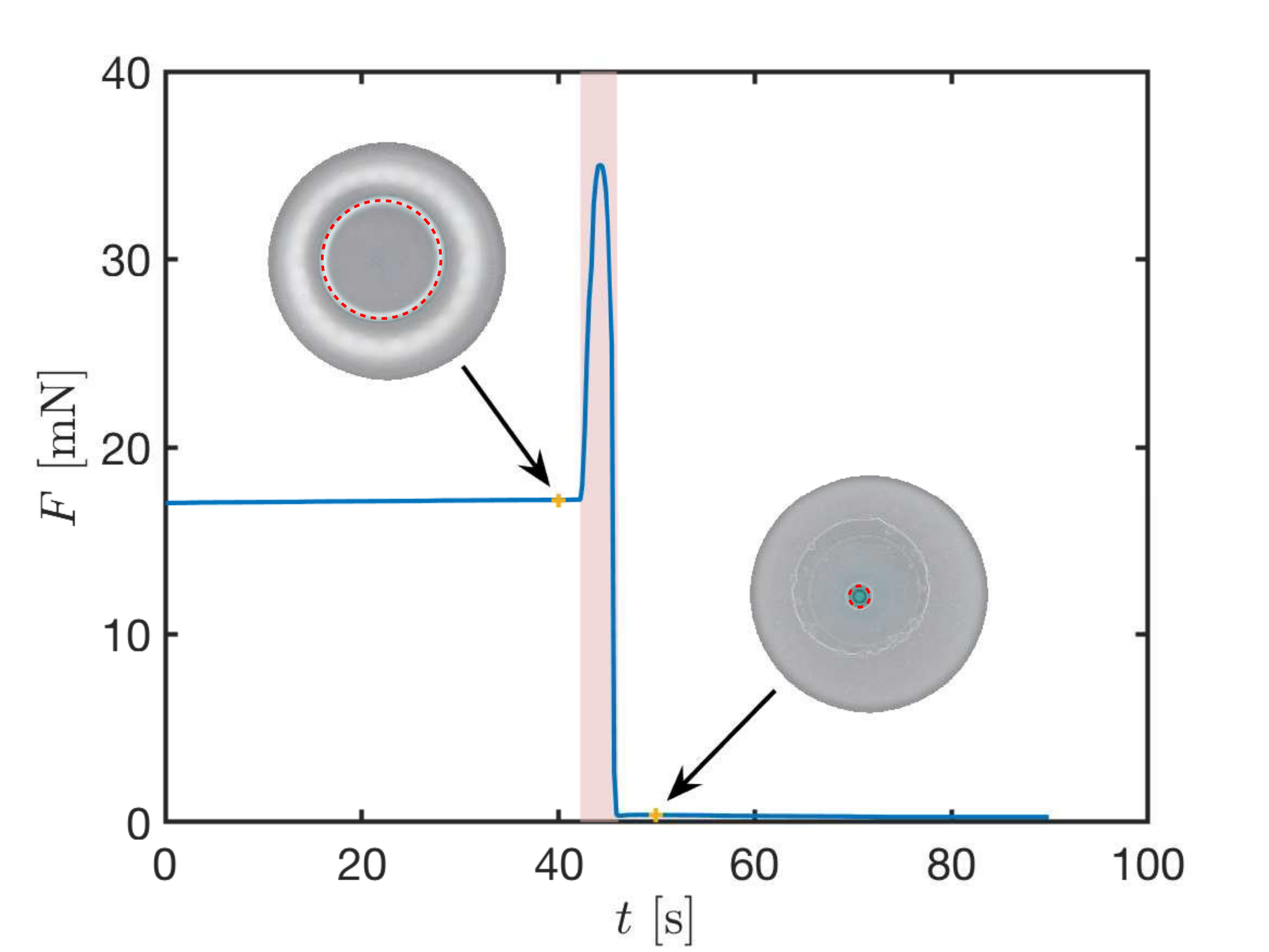}
	\caption{Increasing the tension, while maintaining a fixed separation, results in the `switching off' of strong adhesion. The tension is increased from $T=\SI{2.5}{\newton \per \meter}$ to $T=\SI{7.1}{\newton \per \meter}$ over the duration of the highlighted region, and is constant otherwise. Examples of the droplet spread before and after the tension change are shown, with the meniscus position denoted by a red dashed circle. Here the clamp was fixed at a height $\hinf=\SI{1.1}{\milli \meter}$, with a drop volume $V=\SI{7.8}{\micro \liter}$. }
	\label{fig:TensionChange}
\end{figure}

At fixed gap width, experiments also reveal that when starting in the high-adhesion contacting state it is possible to significantly decrease the adhesion force solely through an increase in the tension (decrease in the deformability $\G$, example shown in fig.~\ref{fig:TensionChange}). This confirms that tension variation might be used as a detachment mechanism. We investigate this possibility further later. 

\subsection{Adhesion testing} \label{sec:AdhesionTest}

\begin{figure}[tbp]
	\centering
	\includegraphics[width=0.29\linewidth]{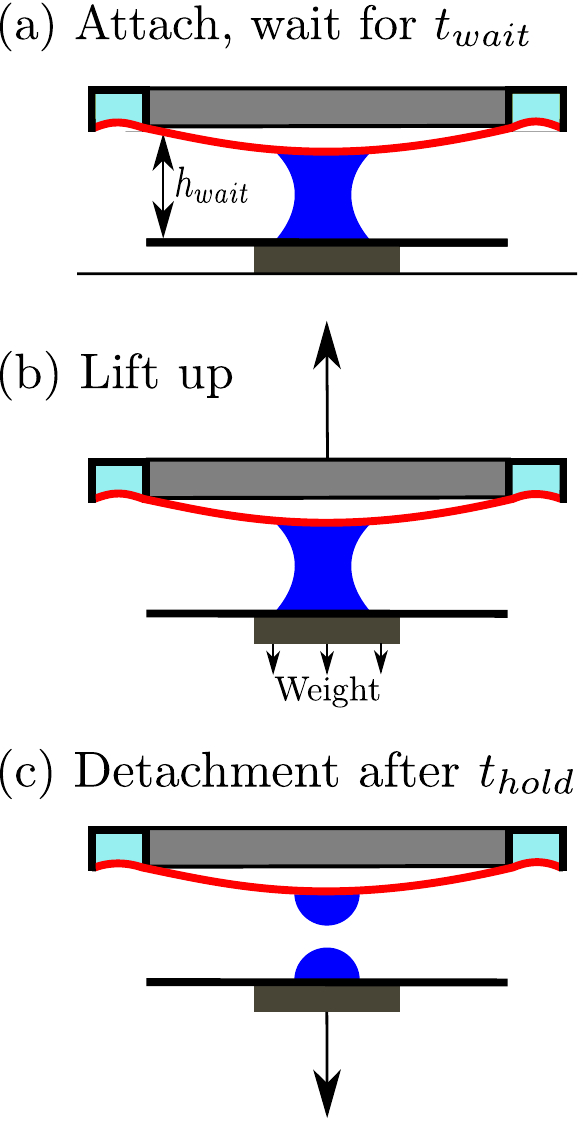}
	\includegraphics[width=0.69\linewidth]{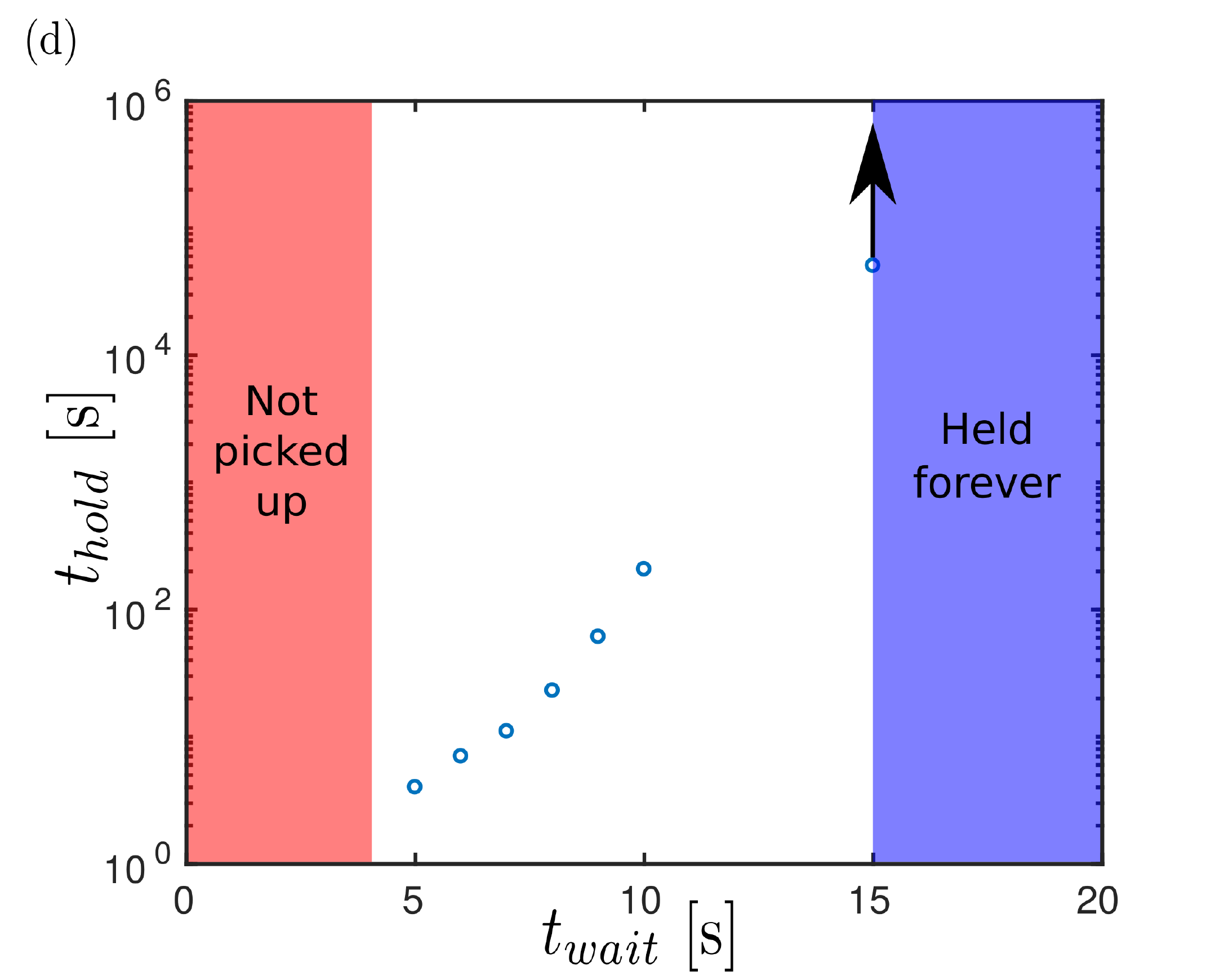}
	\caption{Schematic of the dynamic adhesive test. (a) The clamp is lowered to a set distance from the glass plate. (b) After a time $\twait$ the clamp is raised at a constant speed, lifting the glass plate and load. (c) After some time $\thold$ detachment may occur and the load falls back. 
	(d) Experimental results showing the relationship between the time the system is left to equilibrate, $\twait$, and the adhesion time, $\thold$, for $\hinf = \SI{0.3}{\milli \meter}$, $T = \SI{2}{\newton \per \meter}$, $V = \SI{10}{\micro \liter}$ and lifting a mass of $\SI{2.7}{\gram}$. Experiments were performed with $\twait \leq \SI{4}{\second}$ but no noticeable lift-off was observed. When $\twait \gtrsim 10$\,s the load would remain attached on the time scale of hours, making it difficult to obtain a accurate value for $\thold$, and for $\twait = \SI{15}{\second}$ the load remained attached overnight (over 12 hours).}
	\label{fig:WaitExpt}
\end{figure}

To test the feasibility of this system as an adhesive, we tasked it with lifting some small loads attached to a glass slide (fig.~\ref{fig:WaitExpt}a-c). For the parameters in our tests (${\gamma \cos \theta = \SI{29}{\milli \newton \per \meter}}$,  $T = \SI{2}{\newton \per \meter}$, $V = \SI{10}{\micro \liter}$) the equilibrium theory predicts that a load corresponding to a mass of $\SI{2.7}{\gram}$ can be comfortably supported at a separation $\hinf = \SI{0.3}{\milli \meter}$; moreover, for these parameter values, the equilibrium theory predicts that we should have an equilibrium configuration with the membrane in contact with the glass slide. We note, however, that once a load is lifted then the droplet-membrane system  evolves at fixed force, rather than a given gap separation; we expect either the membrane to be pulled into close contact and remain stuck, or to detach completely.

On lowering the PVS sheet to a separation  ${\hinf = \SI{0.3}{\milli \meter}}$ and lifting immediately, we find that the load drops off. Holding the sheet at this set distance for a short period of time increases the length of time for which the adhesion is successful  (fig.~\ref{fig:WaitExpt}d). Indeed, if we hold the two surfaces at the fixed separation for sufficiently long (on the order of $15\mathrm{~s}$) then the load is adhered indefinitely (timescale of days). This suggests that the dynamics of adhesion are non-trivial and deserve further study. We therefore turn to study the dynamics of adhesion now.

\section{Dynamics} \label{sec:dynamics}

To study the dynamics of adhesion, we use a lubrication-type model: we assume that the flow in the thin gap between the membrane and substrate is viscous, consistent with the assumption of small aspect ratio used in studying the equilibrium of the system. Applying the no-slip boundary condition (i.e.~zero velocity) at both the membrane and wall, the Stokes equation for the flow is readily integrated to give the radial fluid flux as 
\begin{equation} \label{eq:Flux}
	q = -\frac{h^3}{12 \mu} \frac{\partial p}{\partial r}=\frac{T}{12 \mu} h^3 \frac{\partial}{\partial r} (\nabla^2 h), 
\end{equation}
with $\mu$  the dynamic viscosity of the fluid.

Conservation of mass \cite{Leal2007} then gives an evolution equation for the membrane height, $h(r,t)$. This can be non-dimensionalized in the same manner as the static scenarios, i.e.~$H=h/(V/L^2)$ and $R=r/L$, with the natural timescale  ${t_\ast = 12 \mu L^4 / \gamma V \cos \theta}$ used to non-dimensionalize time (note that we denote the dimensionless time by $t$ to avoid confusion with the applied tension $T$). The dimensionless partial differential equation for the evolution of the membrane is then
\begin{equation} \label{eq:PDE}
	\frac{\partial H}{\partial t} = \frac{1}{R} \frac{\partial}{\partial R}  \left[ R H^3 \frac{\partial P}{\partial R}  \right]
\end{equation}
where the dimensionless pressure $P=p/(\gamma L^2 \cos \theta /V)$ is
\begin{equation}
	P = -\frac{1}{\G} \nabla_R^2 H.
\end{equation}

As boundary conditions at the origin ($R=0$) we impose zero membrane slope (axisymmetry) and no radial flux, i.e.
\begin{align}
	\frac{\partial H}{\partial R} &= 0 \qquad \text{at } R=0, 	\label{eq:BCsymm}\\
	\frac{\partial}{\partial R} (\nabla_R^2 H) &= 0 \qquad \text{at } R=0.
\end{align} 
At the  meniscus, $R=R_M$, the pressure in the liquid must match that provided by the pressure jump across the meniscus; this provides a condition on the membrane curvature at $R=R_M$. Also, and as in the static problem, the slope of the membrane and the membrane displacement must both be continuous. Finally,  the membrane must reach its clamped value, $H(1)=\Hinf$. Since the problem in the dry membrane is quasi-static, the membrane shape may be solved analytically in this region for given values of $H_M$ and $R_M$; the result is that there are two conditions at the meniscus, namely
\begin{align}
	\nabla_R^2 H =  \frac{2\G }{ H_M } \qquad \text{ at } R=R_M, \\
	\frac{\partial H}{\partial R}  = \frac{H_M-\Hinf}{R_M \log{R_M}} \qquad \text{ at } R=R_M.
\end{align}

An equation for the motion of the meniscus, $R_M(t)$, is determined by requiring it to have the velocity that balances the flux, i.e.
\begin{equation}
	\frac{\mathrm{d}R_M}{\mathrm{d}t} = - H_M^2 \left. \frac{\partial P}{\partial R} \right|_{R=R_M}.
	\label{eq:BCmeniscus}
\end{equation} (Note that this motion of the meniscus ensures that global conservation of mass, $2 \pi \int_0^{R_M}RH~\mathrm{d}R=1$, is automatically satisfied throughout.)

The  partial differential equation \eqref{eq:PDE} is solved numerically subject to  the boundary conditions \eqref{eq:BCsymm}--\eqref{eq:BCmeniscus} and the initial condition $H(R,0)=\Hinf$. To determine the numerical solution, we discretize space in a flux-conservative manner and evolve in time using the method of lines integrated with MATLAB's ODE solvers. (Further details are given in Appendix \ref{Ap:Numerics}.)

\subsection*{Contacting \& non-contacting dynamics}

\begin{figure}[tbp]
	\centering
	\includegraphics[width=0.85\linewidth]{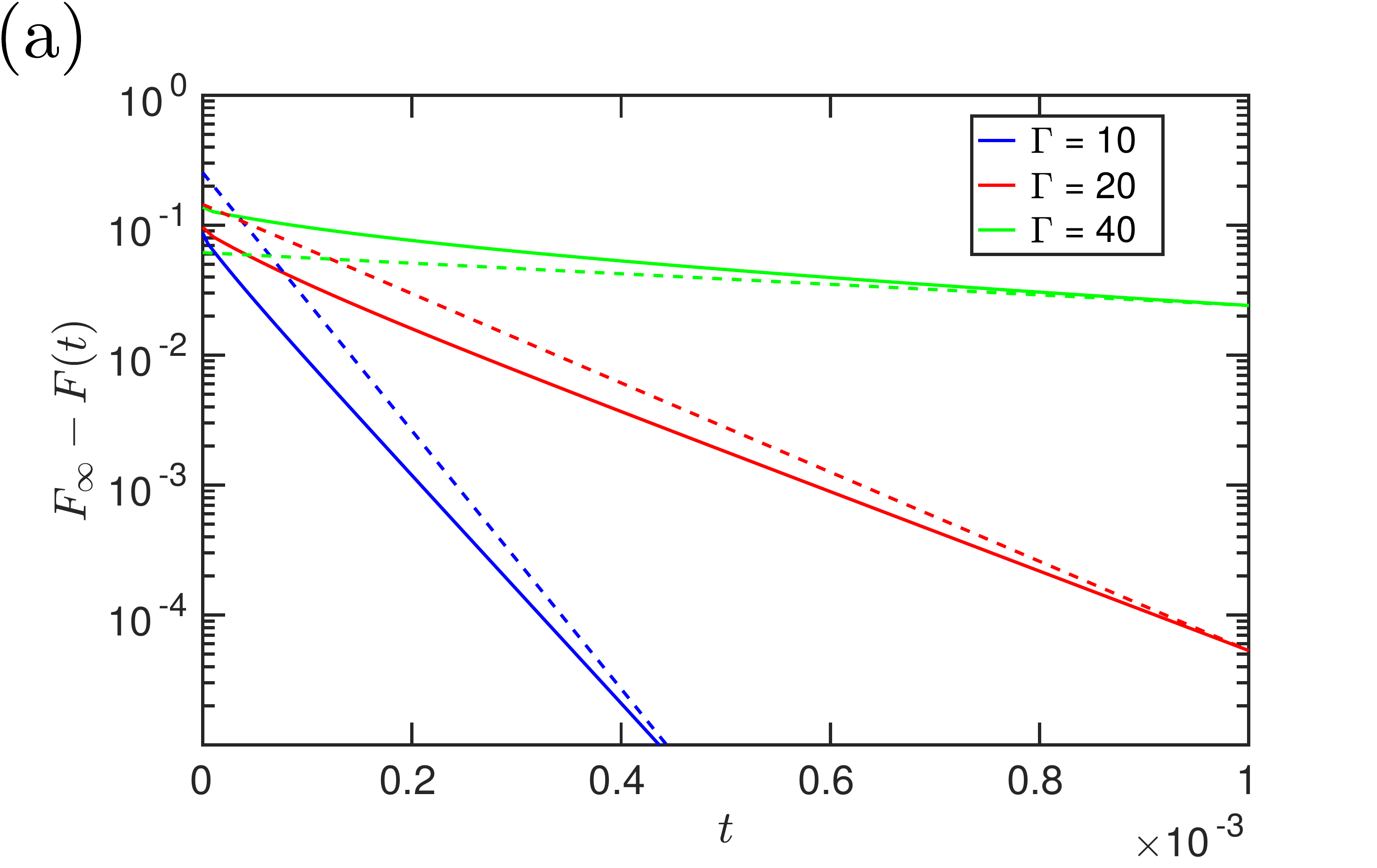}
	\includegraphics[width=0.85\linewidth]{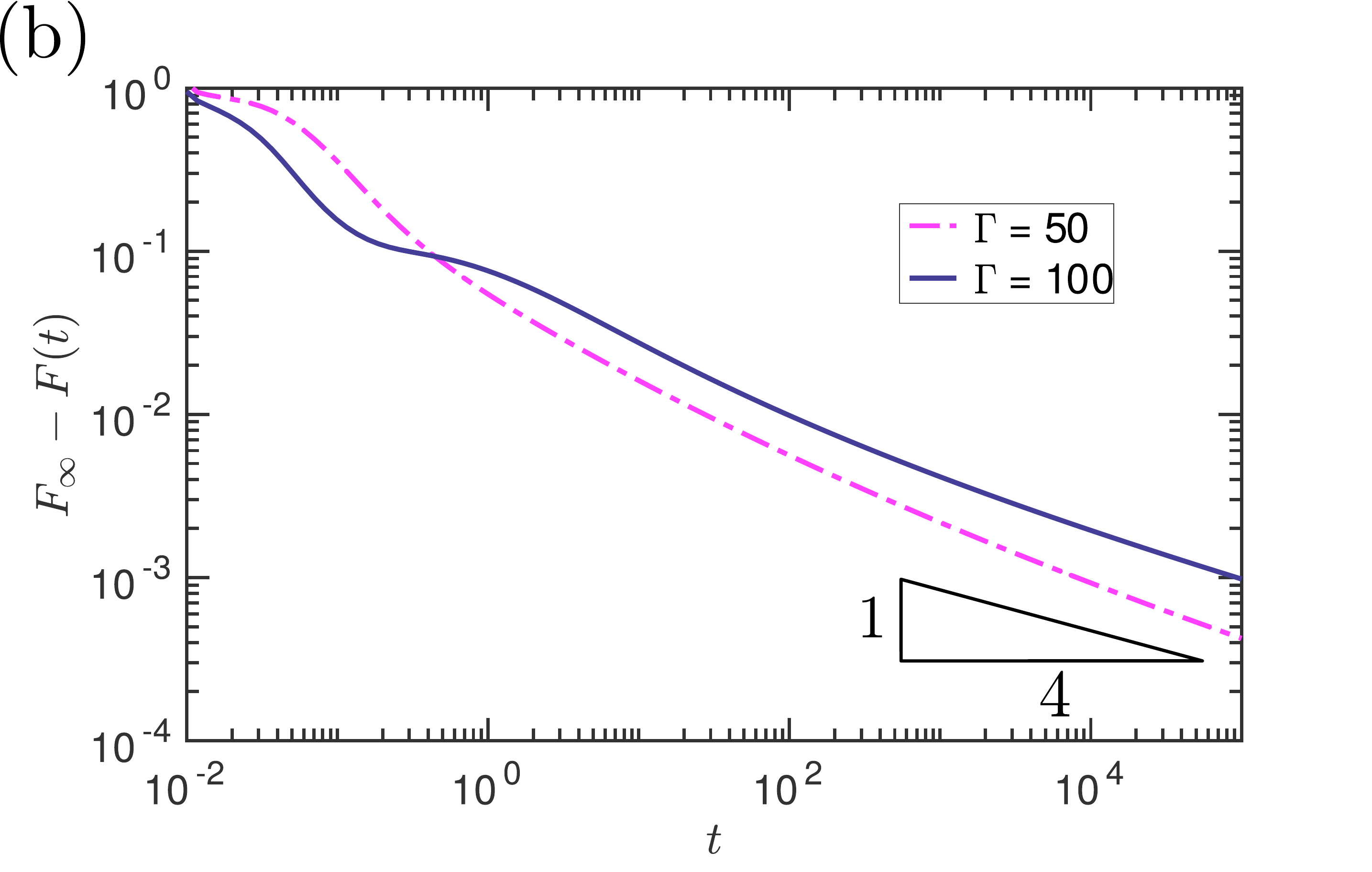}
	\caption{The dynamic approach of the dimensionless adhesive force, $F(t)$, to its equilibrium value $F_\infty$ in both (a) non-contacting and (b) contacting cases. (a) In the non-contacting case, the force  decays exponentially to its equilibrium value, consistent with a linear stability analysis of the equilibrium (dashed lines show the expected decay rate, i.e.~slope, only). (b) In the contacting case, the force instead appears to decay according to a power law. In each case, the equilibrium force, $F_\infty$, is calculated from the static theory, and $\Hinf = 5$.}
	\label{fig:DynamicForce}
\end{figure}

The numerical solutions of the dynamic problem qualitatively confirm the results of the static analysis presented in \S\ref{sec:eql}. In particular, the system appears to have two distinct types of equilibria characterized by whether the membrane and base are in physical contact. Furthermore, it is possible to switch between these states by changing the applied tension, for example. However, the dynamic simulations demonstrate a further key difference between these types of solution: with fixed control parameters ($\G$ and $\Hinf$) the approach to equilibrium is significantly quicker in the non-contacting case than it is in the contacting case (compare the different time scales in figures \ref{fig:DynamicForce}a and \ref{fig:DynamicForce}b). This also agrees qualitatively with our experimental observation in fig.~\ref{fig:Experiment}.

The results of fig.~\ref{fig:DynamicForce} are presented with a fixed gap width $\Hinf$ but with various values of the membrane deformability $\G$. We see that with lower values of $\G$ (more tense, hence less deformable, membranes), the adhesion force decays exponentially to the expected equilibrium value. These  relatively undeformable membranes have equilibria that are out-of-contact; the membrane approaches these equilibria  quickly, and the exponential decay of the adhesive force to the equilibrium value  may be understood by a standard   linear stability analysis about the equilibrium configuration (see Appendix \ref{Ap:LinStab}). Increasing $\G$  to values for which the equilibrium analysis suggests a contacting solution exists, we see (fig.~\ref{fig:DynamicForce}b) that the adhesive force approaches its final value significantly more slowly than might be expected from a linear stability analysis: the decay appears to be power law, rather than exponential. 

\subsection*{Trapped liquid slows contact}

\begin{figure}[tbp]
	\centering
	\includegraphics[width=\linewidth]{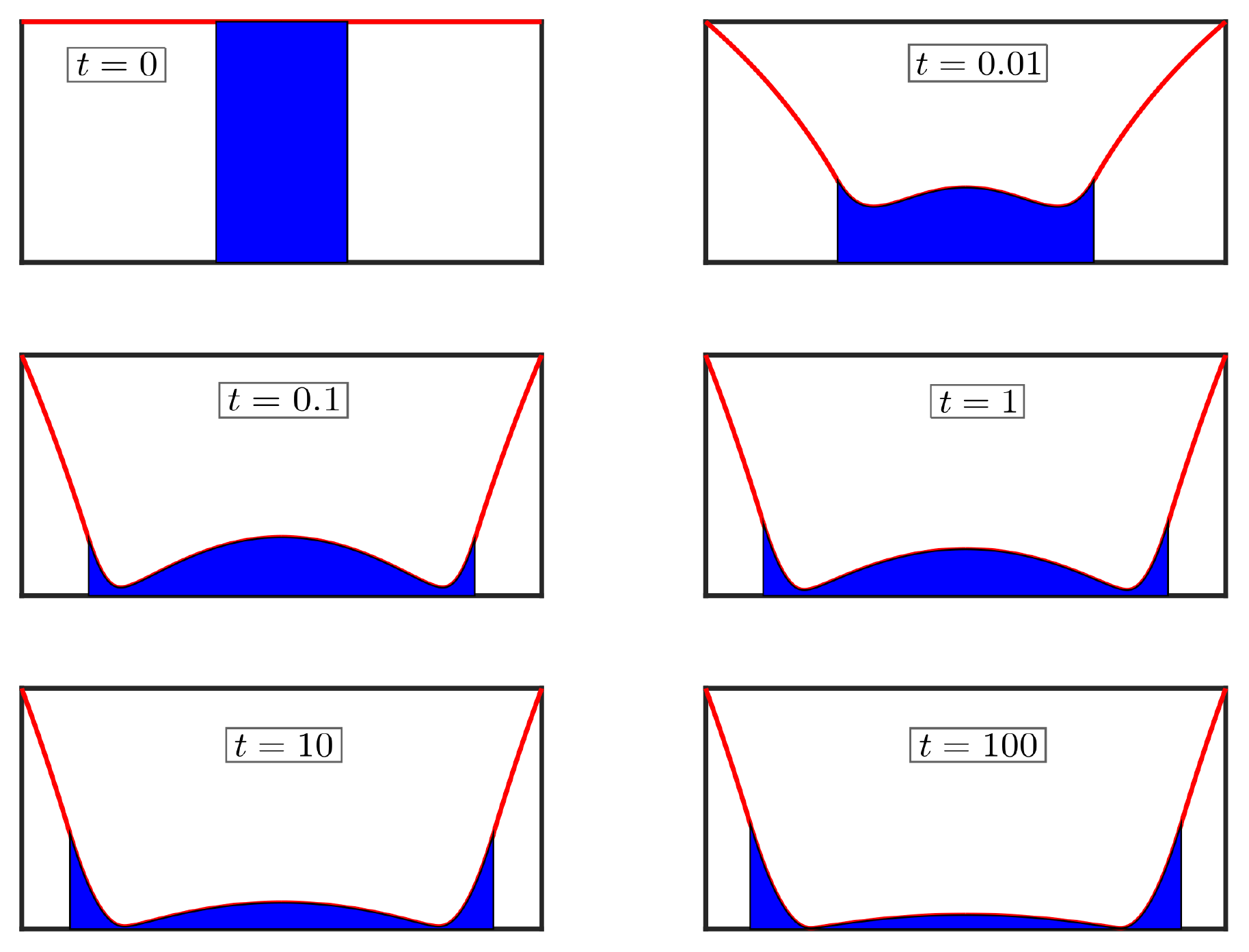}
	\caption{Snapshots of the cross-sectional profile of the membrane and drop during the approach to contact of a highly deformable membrane (here $\G=100, \Hinf=5$). Note that some of the fluid is trapped in a dimple in the membrane and drains very slowly. The initial condition is shown in the top-left panel with the behaviour at five subsequent times also shown. Note that the vertical scale is exaggerated here (by the different scales used to non-dimensionalize horizontal and vertical lengths) --- in reality the drop is  thin and wide.}
	\label{fig:DimpleFormation}
\end{figure}

The different dynamic behaviour in the approach to contact (compared to that out-of-contact) can  be explained by the formation of a fluid dimple under the membrane (see snapshots of the membrane shape from simulations in fig.~\ref{fig:DimpleFormation}). Here the membrane is pulled towards contact by the capillary forces of the droplet, and to accommodate this motion the fluid must be squeezed radially outwards towards the meniscus. However, this flow is sufficiently resisted by viscosity that some of the fluid becomes trapped beneath the membrane and only  drains slowly; we shall see that this slow drainage controls the dynamics at late times.

\begin{figure}[tbp]
	\centering
	\includegraphics[width=\linewidth]{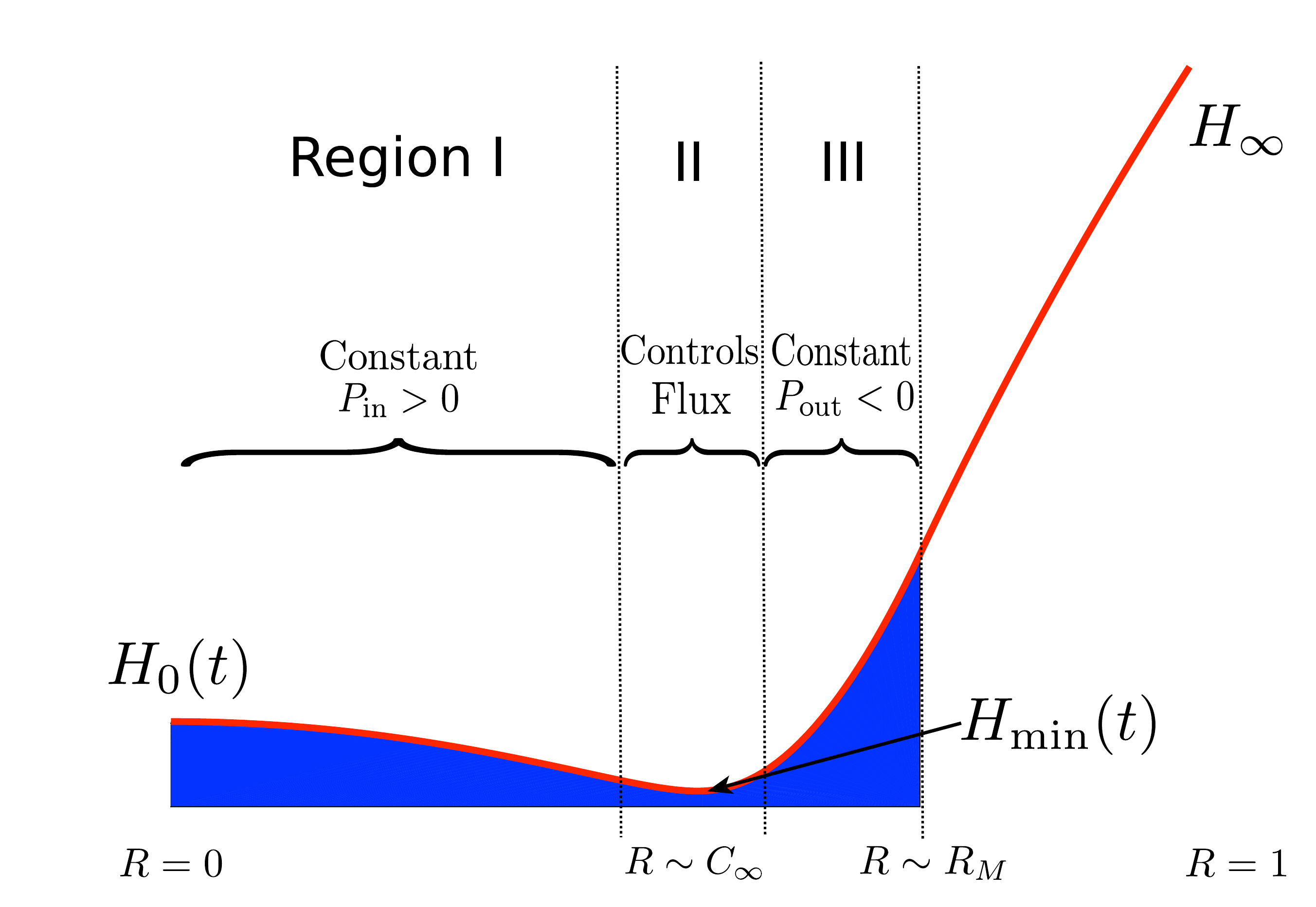}
	\caption{When approaching contact, a dimple forms beneath the membrane. To understand the evolution of the dimple, we split the fluid into 3 distinct regions: region I is the dimple at uniform positive pressure with dimple height $H_0(t)$, region III is the meniscus region at uniform negative pressure, and region II is a small annular region surrounding the narrow gap, $\Hmin$, that controls the fluid flux, and is at a radial position close to the edge of the equilibrium contact region, $C_\infty$.}
	\label{fig:DimpleDiagram}
\end{figure}

The dimple formation shown in fig.~\ref{fig:DimpleFormation} is reminiscent of previous work by Jones \& Wilson \cite{Jones1978} and Yiantsios \& Davis \cite{ Yiantsios1990} on a  bubble  approaching an interface or rigid wall through a viscous liquid. We use these studies as templates with which to study the dynamics of our system.  At sufficiently late times the bulk of the fluid inside the dimple is at uniform pressure, and likewise for the fluid outside. Crucially, however, the two regions have different pressures and are joined by a narrow region at the dimple edge, in which the shape of the membrane controls the leakage flux; the structure of the problem is shown in fig.~\ref{fig:DimpleDiagram}.

In fig.~\ref{fig:DimpleDiagram}, we see that the portion of the membrane that separates the regions of approximately constant pressure is narrow. We expect that, once the dimple has formed, this narrow gap will be located close to the equilibrium contact point, $C_\infty$, calculated from the equilibrium theory and so take its location to be at $C_\infty$, to leading order. We assume that the height there, $\Hmin=H (C_\infty)$, is negligible in comparison to the height away from $R=C_\infty$. With this assumption, we see from Reynolds' equation for the dimensionless flux, $ Q = - H^3 \partial P/\partial R $, that the pressure gradient is largest where the gap width $H$ is small,  justifying taking the pressure to be uniform away from $R=C_\infty$.

Inside the dimple the uniform  pressure is positive and dictates the membrane shape via a Poisson equation. The profile of the dimple may therefore be written ${H(R) \approx H_0 (1- R^2/C_\infty^2)}$ where $H_0$ is the height at the centre ($R=0$). Integrating, we find the volume trapped within the dimple is $V_{\text{dimple}} \approx \pi C_\infty^2 H_0/2$.

Outside the dimple the constant fluid pressure must be negative (since it must match the pressure at the meniscus); we therefore write  this pressure $P=-\Pi$  with $\Pi>0$. As with the dimple region, the  flow here is negligible, so that the membrane shape evolves quasi-statically (though for simplicity we do not give the shape of the membrane in this region explicitly here).

In the small gap at the edge of the dimple (which we call the `neck', illustrated by region II in fig.~\ref{fig:DimpleDiagram}) the volumetric flux of fluid, $Q$, is controlled by the local membrane geometry via a lubrication flow. We introduce a local coordinate $X = R-C_\infty$ in the neck region (i.e.~$|X|\ll C_\infty$) so that
\beq
Q \approx \G^{-1} H^3 \frac{\partial^3 H}{\partial X^3}
\label{eq:LocalFlux}
\eeq here.

Equation \eqref{eq:LocalFlux} can be solved to give the shape of the membrane in the neck region, but requires matching conditions as $X\to\pm\infty$, as well as a value for the flux $q$. In these local coordinates  the curvature due to the pressure in the meniscus requires ${\partial^2 H/\partial X^2 \to \G \Pi}$ as ${X \to \infty}$; similarly, the membrane gradient at the  interior edge of the neck region requires ${\partial H/ \partial X \to -2H_0/C_\infty}$ as ${X \to -\infty}$. Finally, the integrated flux through the neck region, $2 \pi C_\infty Q$, must balance the rate at which the volume of the dimple decreases, $\dot{V}_{\text{dimple}}$. Combining these three relations with \eqref{eq:LocalFlux} we obtain the scaling relations  
\[		\frac{\Hmin}{X_\ast^2} \sim \G \Pi,		\qquad		\frac{\Hmin}{X_\ast} \sim \frac{H_0}{C_\infty},		\qquad		\frac{C_\infty^2 H_0}{t} \sim \frac{C_\infty \Hmin^4}{\G X_\ast^3}	\]
where $X_\ast$ is the typical horizontal scale in the neck region. Solving these equations leads to leading order scalings for the dimple height $H_0$, as well as the  height $\Hmin$ and width $X_\ast$ in the narrow gap region, in terms of $\G, \Pi, C_\infty, t$ as follows:
\begin{equation} \label{eq:DimpleScaling}
\begin{aligned}
		H_0 &\sim \G^{1/2} \Pi^{1/4} ~C_\infty^{3/2} ~t^{-1/4}, \\
		X_\ast &\sim \G^{-1/2} \Pi^{-3/4} ~C_\infty^{1/2} ~t^{-1/4}, \\
		\Hmin &\sim \Pi^{-1/2} ~C_\infty ~t^{-1/2}.
\end{aligned}
\end{equation}

This scaling analysis reveals that the dimple height $H_0 \sim t^{-1/4}$, and the membrane height in the narrow gap $\Hmin\sim t^{-1/2}$, similar to the scalings obtained by Jones \& Wilson \cite{Jones1978} in drop coalescence. Thus contact between the membrane and base will not occur in finite time (unless another shorter-range force, such as van der Waals, takes over). 

We also find that the similarity solution for the membrane shape in this transition region is the same as that found by Jones \& Wilson \cite{Jones1978} (see Appendix \ref{Ap:SimSoln}). However, the pre-factors in the scaling relations \eqref{eq:DimpleScaling} differ because in this problem the outer curvature is set by the pressure at the meniscus (i.e.~the meniscus height) rather than, for example, a bubble radius or volume. These pre-factors may be found in terms of the membrane deformability  $\G$ and the equilibrium contact point $C_\infty$ (which is itself a function of $\G$ and $\Hinf$); the pre-factors also depend on the meniscus pressure $\Pi$ (at late times, we expect that $\Pi \sim 2/H_M^\infty$, with $H_M^\infty$ the equilibrium meniscus height, since the pressure is set by the meniscus curvature).
 
Our main focus here is on the evolution of the key properties of the system at late times, especially the adhesion force (but also the meniscus position and height). To  progress, we assume that the membrane behaves quasi-statically outside the dimple, evolving due to the volume increase as fluid leaks  through the neck region. The meniscus position, $R_M$, its height, $H_M$, and the (effective) contact position, $C$, therefore obey \eqref{eq:ContactCondition}--\eqref{eq:ContactVol}, but with the left hand side of \eqref{eq:ContactVol} modified to $1-V_{\text{dimple}}$ to account for the (decreasing) amount of  fluid trapped within the dimple. We note that (see Appendix \ref{Ap:SimSoln})
\begin{equation}
	V_{\text{dimple}} \sim \frac{\pi}{2} A \G^{1/2} \Pi^{1/4} C_\infty^{7/2} t^{-1/4} \label{eq:DimpleVol}
\end{equation}
for a constant $A \approx 0.20$ that is found numerically. (The radial flux is then $Q \sim \dot{V}_{\text{dimple}}/2 \pi C_\infty \sim t^{-5/4}$.)

Expanding $R_M$, $H_M$ and $C$ about their equilibrium values, and using the  leading order expression for dimple volume, \eqref{eq:DimpleVol}, we  linearize these three conditions to calculate their first order corrections (Appendix \ref{Ap:Force Evolution}). The correction to the adhesion force is then determined from linearizing the relation
\begin{equation}
	F = -2 \pi \int_C^{R_M} R P(R,t) ~\mathrm{d}R = \frac{2 \pi}{\G} \frac{ H_M - \Hinf}{\log R_M}
\end{equation}
which leads to
\beq
F \sim F_\infty - F_1 ~t^{-1/4}
\label{eq:algFdecay}
\eeq
for $F_1$ a constant that can be computed numerically (see Appendix \ref{Ap:Force Evolution}). 
Note that here $F_\infty$ is the force calculated from the static theory; although the pressure within the dimple is positive (and hence slightly reduces the adhesive force), at late times this correction is small and the adhesive force is dominated by the suction pressure outside the dimple.

\begin{figure}[tbp]
	\centering
	\includegraphics[width=\linewidth]{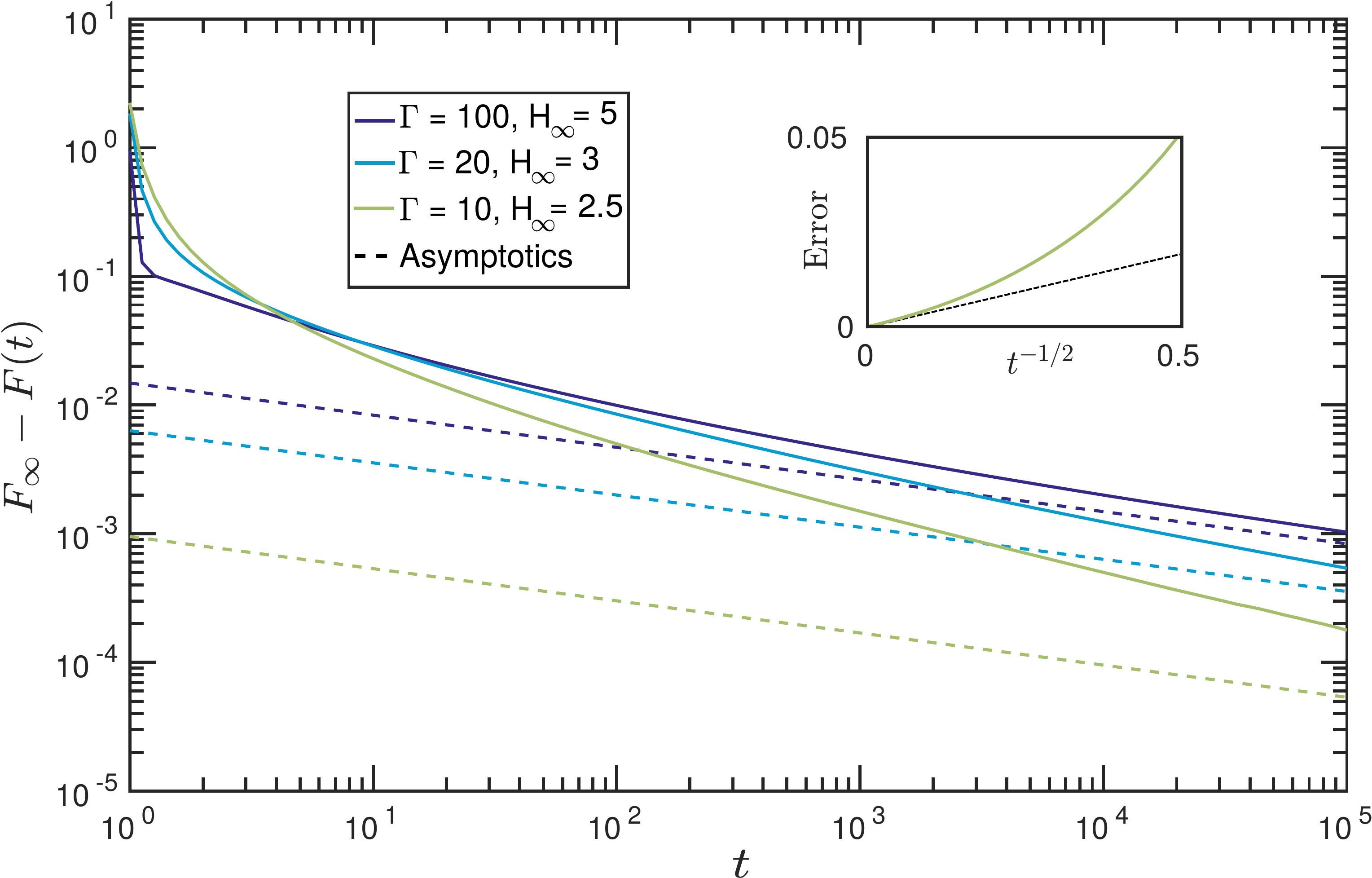}
	\caption{The adhesion force decaying to its equilibrium value, $F_\infty$ (calculated from the analysis of \S\ref{sec:eql}), in the contacting regime. The numerics (solid curves) approach the asymptotic dimple theory (dashed lines) at late times. Inset: the absolute error (solid curve) between the numerically calculated force $F(t)$ and the first two terms of the asymptotic expansion is approximately linear (black line for comparison) in $t^{-1/2}$ at late times (here  $\G=10, \Hinf=2.5$).  }
	\label{fig:ForceDynamics}
\end{figure}

Fig.~\ref{fig:ForceDynamics} shows  the numerically-determined decay of the adhesion force  to its equilibrium value, together with  the full prediction of the asymptotic theory (including pre-factors). Qualitatively we see a reasonable match between the two quantities at late times, suggesting that the expected algebraic decay of \eqref{eq:algFdecay} is indeed that observed numerically. The inset of fig.~\ref{fig:ForceDynamics}  shows the absolute error in \eqref{eq:algFdecay} and confirms that the error does indeed occur at higher order (i.e.~at $O(t^{-1/2})$). 

While the leading-order asymptotic results are in good agreement with numerical simulations for very late times, the convergence is relatively slow: our results suggest an expansion in powers of $t^{-1/4}$ so that to obtain even a $10\%$ error requires $t=O(10^4)$ or greater. The numerical results shown in fig.~\ref{fig:ForceDynamics} confirm that the $t^{-1/4}$ scaling is only observed for dimensionless times ${t\gtrsim10^4}$. In our experiments, the dimensional time scale ${t_\ast = 12 \mu L^4 / \gamma V \cos \theta \approx \SI{400}{\second}}$; the time needed to observe this dynamic scaling with our experimental parameters would therefore be on the order of weeks, and effectively not observable.

\section{Detachment} \label{sec:detach}

Our equilibrium theory has shown that, from the perspective of maximizing the adhesive force, it is beneficial to be in the contacting regime. Contact can be achieved either by decreasing the gap width (decreasing $\Hinf$) until the membrane snaps to contact or by decreasing the tension sufficiently (increasing $\G$). As shown in the last section, the larger forces associated with contact are moderated by the caveat that the force approaches its higher contacting value relatively slowly,  $F_\infty-F(t)\sim t^{-1/4}$. Nevertheless, the order of magnitude increase in adhesion force seen as contact is approached, and the reasonable waiting times for significant attachment observed in \S\ref{sec:AdhesionTest}, suggest that operating in the $(\G,\Hinf)$ parameter regime corresponding to contacting equilibria is still beneficial.

Having investigated adhesion to a substrate, and seen the importance of being close to contact, it is natural to then ask how can one detach from the surface efficiently? In particular, if one begins close to contact, is there a `best' way to unstick? The key quantity of interest is the effective work of separation, which we define to be the work done to separate the surfaces, 
\beq
\Wadh=\int_{\Hinf^0}^\infty F~\mathrm{d} \Hinf+\Delta U_{\mathrm{elast}},
\label{eq:Wadh}
\eeq 
where $\Hinf^0=\Hinf(t=0)$ is the initial gap width and $\Delta U_{\mathrm{elast}}$ is the change in elastic energy due to stretching of the sheet. Note that the upper limit of integration in \eqref{eq:Wadh} is $\Hinf=\infty$, since we want to completely separate the surfaces. 

In practice we calculate the work of separation in the quasi-static case that follows from the increase in the total equilibrium energy of the sheet. Here the energy input will be stored in two different ways: elastic energy in the stretched sheet and surface energy at the interfaces. Ignoring small terms the dimensionless energy, $U$, can be written as 
\beq \label{eq:QSEnergy}
\G U = \pi \int_0^1 R \left( \frac{\mathrm{d}H}{\mathrm{d}R} \right)^2 \mathrm{d}R - 2 \pi \G  R_M^2.
\eeq
In dynamic scenarios, we focus on the case where $\Delta U_{\mathrm{elast}}=0$ and so omit a thorough definition of this term here.

\subsection*{Quasi-static detachment}

We approach the problem of detachment by considering an initial condition corresponding to a contacting equilibrium state (i.e.~strong adhesion). The detachment problem is then to choose a path in ($\G,\Hinf$)-space that minimizes the work of separation, $\Wadh$. 

\begin{figure}[tbp]
	\centering
	\includegraphics[width=\linewidth]{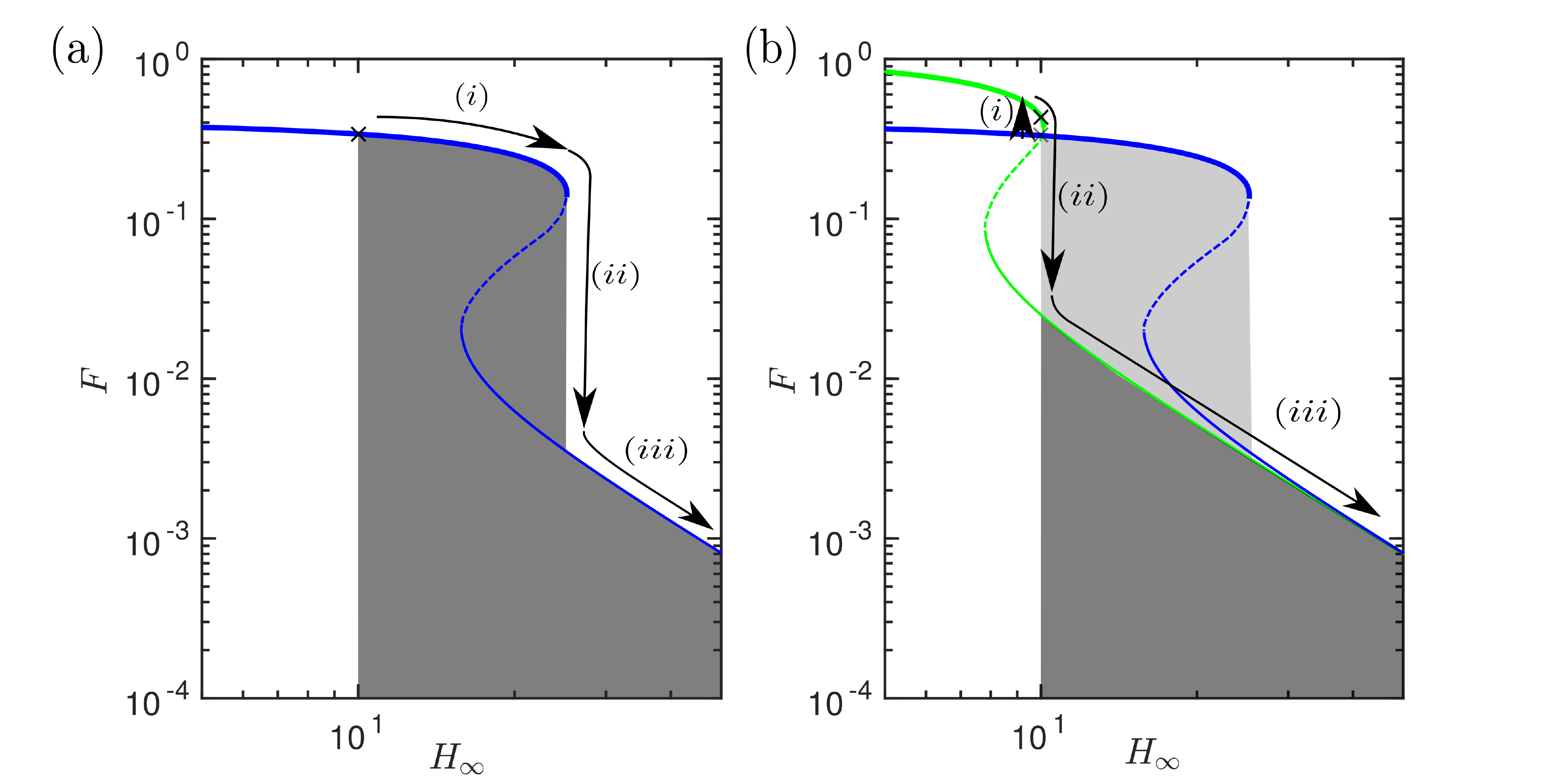}
	\caption{Work of separation from different methods of detachment.
	(a) Pull directly away from the surface (`yanking'), $\Hinf \to \infty$ with $\G$ fixed. The solution follows the equilibrium value (blue solid curves), and jumps from contact to non-contact at the fold (ii). The work of separation is the integral $\int F ~\mathrm{d} \Hinf$, and is illustrated by the shaded dark grey area. 
	(b) If we instead first increase the tension (decrease the deformability $\G$) at fixed $\Hinf$  and then yank, we expect that the work of separation may be significantly reduced. The different stages are: (i) Initially decrease $\G$ (to green solution) before (ii) increasing $\Hinf$ to pull away through the snap-off transition. The total work is the sum of dark grey area and the work done to increase the tension in stage (i) and will be preferable to yanking provided that the energy needed to stretch the membrane in (b)(i) is less than the yanking energy saving compared to (a), illustrated by the light grey region.}
	\label{fig:WorkofAdhesion}
\end{figure}

A simple option is to pull directly away from the substrate at fixed tension: we term this `yanking', and illustrate this path, together with the associated work of separation, in fig.~\ref{fig:WorkofAdhesion}a. This requires working directly against the strong adhesion force of contact. However, we may also vary the tension in the membrane and so another possibility is to increase the tension (decrease $\G$) whilst keeping $\Hinf$ fixed (at least initially). We have already seen experimentally that decreasing $\G$ sufficiently results in the membrane `peeling' off the base (fig.~\ref{fig:TensionChange}), losing contact at a smaller gap separation $\Hinf$ , i.e.~without first pulling the membrane up. Once out of contact, the adhesion force is smaller and so we might expect to then be able to increase $\Hinf$ with significantly less resistance than if yanking without a tension change. This alternative mode of detachment is shown in fig.~\ref{fig:WorkofAdhesion}b, together with a cartoon of how much energy might be saved in this way. Of course, decreasing $\G$ itself  has an energetic cost, $\Delta U_{\mathrm{elast}}$,  and so we must consider the trade-off between the work required to increase the tension and the subsequent reduction in work done against adhesion due to this tension change.

From a quasi-static perspective, the key piece of information is where in the parameter space the path crosses the discontinuous jump from contact to non-contact. As this discontinuity is passed, the system will lose energy that cannot be regained. We therefore need to consider separately the energy change, calculated using \eqref{eq:QSEnergy}, both before and after this jump to determine the energy required to detach.

\begin{figure}[tbp]
	\centering
	\includegraphics[width=\linewidth]{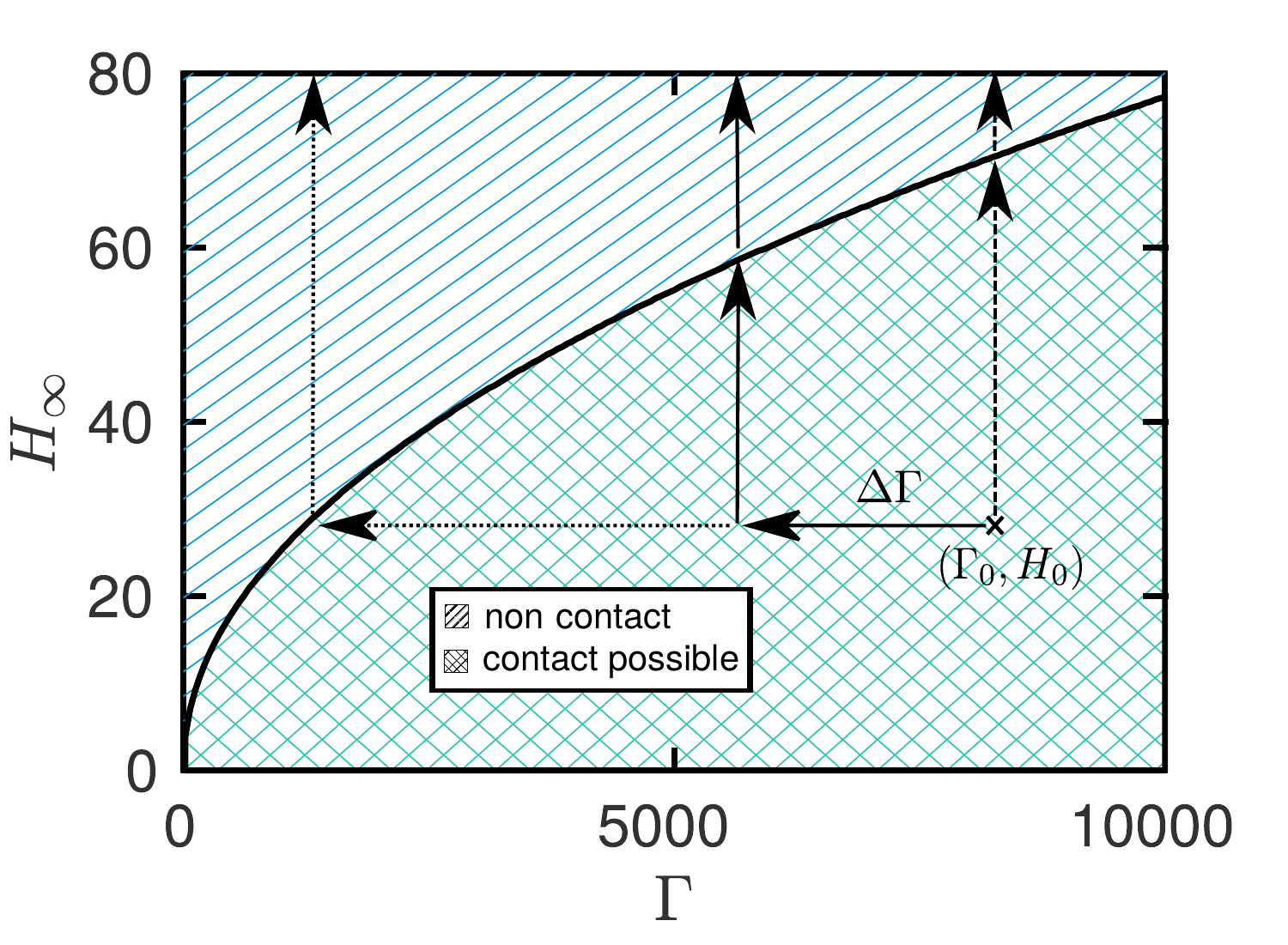}
	\caption{Starting from an initial contacting state $(\G_0,H_0)$, there are many possible paths in parameter space that result in detachment. The simplest of these is a direct pull off at constant tension (`yanking', dashed path). We investigate various quasi-static paths that involve decreasing $\G$ by $\Delta \G$  before taking $\Hinf \to \infty$ (solid path); we find that the optimal (smallest work of separation), is to decrease $\G$ until contact is lost before pulling away (dotted path). Note that here the area labelled `contact possible' contains the three solution region seen in fig.~\ref{fig:Multiple Solns}, and we have omitted the region with no solutions for simplicity.}
	\label{fig:DetachPaths}
\end{figure}

We consider paths where  $\G$ is initially decreased (the tension is increased) by an amount $\Delta \G$, before pulling away ($\Hinf \to \infty$) as illustrated in fig.~\ref{fig:DetachPaths}. $\Wadh$ is calculated from the change in surface and stretching energy of the sheet. We find that increasing the tension results in an overall energy saving. In fact, the best strategy is to increase the tension until contact is lost (at which point the adhesion force is substantially lower) before pulling away (increasing $\Hinf$). Surprisingly, however, the benefit of this change is relatively modest: in simulations the reduction in $\Wadh$ made by increasing the tension first was typically in the region of 5--10\%. While this is surprising, we must also consider the effect of the \emph{rate} of yanking on the work of separation:  since we saw that the dynamics of adhesion significantly modify the equilibrium picture, it is natural to wonder whether the same might be true of dynamic detachment.

\subsection*{Dynamic detachment}

\begin{figure}[tbp]
	\centering
	\includegraphics[width=0.9\linewidth]{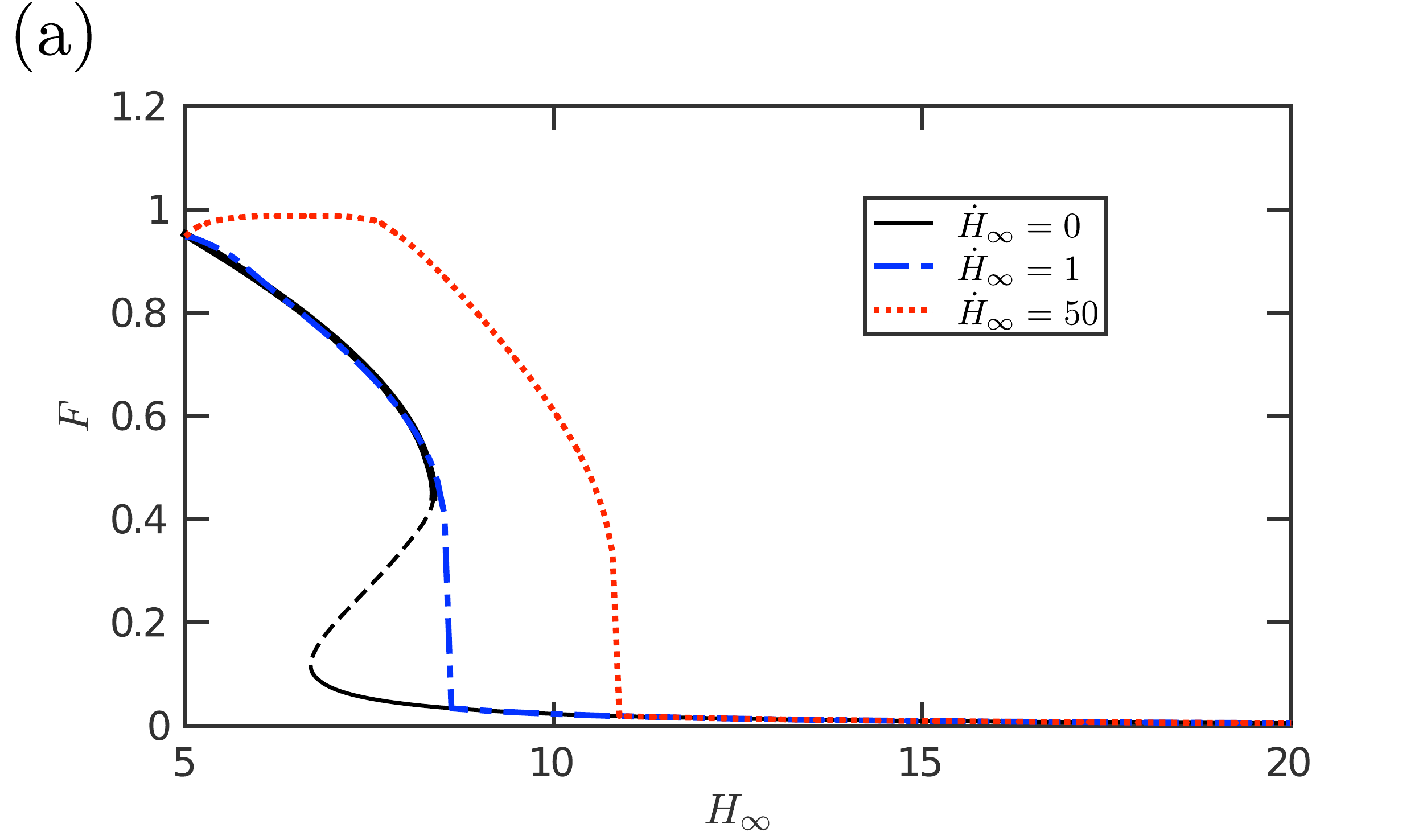}
	\includegraphics[width=0.9\linewidth]{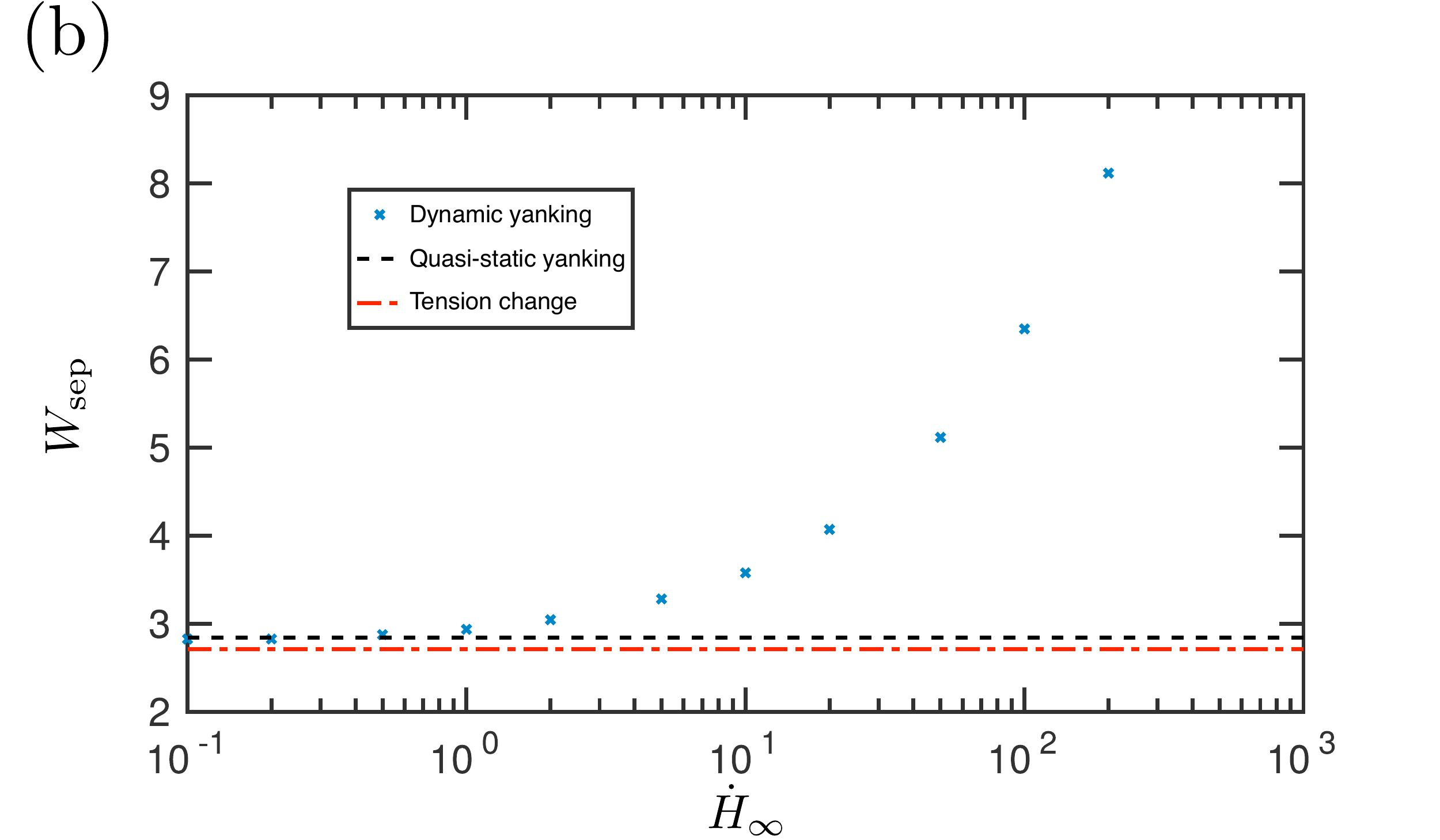}
	\caption{(a) Pulling the membrane off at a faster rate results in a larger adhesion force over a wider range of gap widths, $\Hinf$. (b) The energy required to detach ($\Hinf \to \infty$) when retracting at constant speed increases with the retraction speed, $\dot{H}_\infty$. Dynamic simulations (blue `x') are performed at different constant retraction rates and compared to the quasi-static change in energy at fixed $\G$ (black dashed) and the optimal quasi-static strategy (red dash-dotted) in which $\G$ is decreased until contact is lost before pulling away at fixed $\G$. In both figures the initial condition for dynamic simulations is close to the contacting equilibrium with parameters $\G=100, \Hinf=5$. }
	\label{fig:DynamicEnergy}
\end{figure}

We performed numerical simulations in which the edge height $\Hinf$ is increased at a constant pulling rate, $\dot{H}_\infty$, while $\G$ is maintained at a constant value.
This  shows that the instantaneous adhesive force at a given edge height has a significant rate dependence (fig.~\ref{fig:DynamicEnergy}a): at high rates, viscous forces become important and resist the separation of the membrane from the substrate, which is the same mechanism as for so-called `Stefan adhesion' \cite{Labonte2015,Bikerman1968,Brau2016}. The peak adhesion force is increased, and the force remains high over a larger range of gap widths $\Hinf$. We therefore expect that the work of separation in this case, ${\Wadh = \int F ~\mathrm{d}\Hinf}$, will be significantly increased. An illustration of how the work of separation increases with the rate of detachment is shown in fig.~\ref{fig:DynamicEnergy}b for a particular choice of $\G$ and $\Hinf(t=0)$. Details of the numerical calculation of $\Wadh$ are given in Appendix \ref{Ap:Dynamic Work Calc}. In this example, the work done against the adhesion increases markedly when $\dot{H}_\infty\gtrsim10$ and by more than a factor of two between $\dot{H}_\infty=1$ and $\dot{H}_\infty=100$.

We note that we compare dynamic yanking with the optimal quasi-static strategy (changing the tension in the membrane); this quasi-static model does not include dynamic effects for several reasons. Firstly, the stretching sheet always has a uniform tension, and this tension therefore cannot do any work against the viscous shear force, and there is no rate-dependence in the varying tension model. (However, we note that in this scenario, the dynamics of detachment may be non-trivial, as suggested by detailed studies of `peeling by pulling' \cite{Mcewan1966, Lister2013}.) Secondly, although one might expect rate to have an effect once  contact between the membrane and base is lost and we switch to yanking, the rapid decay towards equilibrium and the small adhesion force suggests that this contribution is negligible in determining the work of separation.

We conclude that it is energetically favourable to detach at a slow speed when only `yanking'. However, in  applications for which detachment is required to occur within a shorter time scale, slow pulling rates are impractical: in our experiments the relevant time scale is $\SI{400}{\second}$. In such scenarios, detachment within seconds would require a significant $\dot{H}_\infty$ and hence a non-negligible increase in the adhesion force (and work of separation) compared to the quasi-static case. This means that the optimal strategy of first increasing the tension could have a substantial benefit over `yanking' in scenarios such as our laboratory experiments; it may also be that this is a preferential method for detaching quickly in other scenarios.

\section{Discussion and conclusions} \label{sec:concl}

Using a simple theoretical model of a tense membrane that is attached to a rigid substrate by a fluid droplet, we have studied some key features of elasto-capillary adhesion. Our model system highlights that deformability can improve adhesive capabilities, with the coupling between the surface tension and elasticity resulting in an increase in the adhesion force over the comparable rigid case. As in similar scenarios \cite{Duprat2015,Wexler2014}, the capillary forces can deform surfaces sufficiently that they  form solid-solid contact. This contact can result in stronger adhesive forces, even without accounting for additional adhesion between the two solid surfaces. 

Two key parameters determine the equilibrium behaviour of our system: the dimensionless gap separation, $\Hinf$, and the membrane deformability, $\G$. These can be actively (and separately) controlled via the clamp height, $\hinf \propto \Hinf$, and the imposed membrane tension, $T \propto \G^{-1}$, respectively. Crucially, we show that by a careful choice of these parameters we can switch the strong adhesion of contact on or off. However the dynamic approach to contact is found to be relatively slow: the adhesion force obeys a scaling law $F(t) - F_\infty  \sim t^{-1/4}$ at late times because fluid is trapped in a dimple and must be squeezed outwards through a narrow gap to allow contact. This slow squeezing out of fluid complicates the use of this adhesive to lift loads:  adhesion is not maintained if the surfaces are not left together for a sufficient period of time. However, a quantitative understanding of this effect is still lacking since, in practice, the adhesion fails via peeling from one edge, which is a feature that cannot be reproduced by the axisymmetric theory presented here. 

Active control of the two parameters $\Hinf$ and $\G$ allows us to develop different strategies to adhere and detach effectively from the substrate, and we suggest that, from an energetic perspective, it is better to `peel' off the base by increasing the tension than to `yank' by pulling directly against the strong adhesion.

The relatively simple framework of this theory has allowed the exploration of some of the key features of adhesion in these systems. 
However, our analysis neglects some important aspects of surface deformation. For example, we require that the membrane is in close proximity to the substrate so that membrane slopes are small and the droplet is sufficiently squashed that we can ignore the effects of the circumferential meniscus curvature and the capillary line force. We also assume for simplicity that the tension in the membrane dominates  bending, additional stretching in the membrane, and the surface tension acting at the meniscus.  As well as understanding the effect of lifting these restrictions, it would be interesting to see how adding extra effects, such as solid-solid adhesion or surface roughness, would modify our results. 
A crucial feature  that is present in insects, but that is missing from our adhesive device, is significant adhesion in the presence of shear \cite{Dirks2011}.

\acknowledgments

This research was partially supported by   the European Research Council under the European Horizon 2020 Programme, ERC Grant Agreement no. 637334 (DV). We have benefited from discussions with J.~R.~Lister.

\appendix

\section{Numerical scheme} \label{Ap:Numerics}

The free boundary problem \eqref{eq:PDE}--\eqref{eq:BCmeniscus} is transformed to a fixed domain problem using the change of variables ${\xi = R/R_M}$. Defining $\alpha(\xi)=R_M^2 H$ then allows us to re-write the governing PDE in the form of a conservation equation $\partial \alpha / \partial t + \nabla \cdot \mathbf{\hat{Q}} = 0$, where the flux $\mathbf{\hat{Q}}=\hat{Q}\mathbf{e_\xi}$ is given by
\begin{equation}
	\hat{Q} = -\frac{\alpha^3}{R_M^6}  \frac{\partial \hat{P}}{\partial \xi} -  \frac{ \xi \alpha\dot{R}_M}{R_M}, \qquad \hat{P} = \frac{-1}{\G R_M^4} \nabla_\xi^2 \alpha,		
\end{equation}
with the meniscus position evolving according to the equation
\begin{equation}
\dot{R}_M = - \frac{\alpha(1)^2}{R_M^5} \left. \frac{\partial \hat{P}}{\partial \xi}  \right|_{\xi=1}.
\end{equation}

We solve this system numerically using the method of lines. The domain $\xi \in [0,1]$ is split into concentric annuli ${\xi_{i-1/2}<\xi<\xi_{i+1/2}}$ with uniform height $\alpha_i$ and pressure $\hat{P}_i$ ($i=1,2, \dots n$) where $\xi_j=j/n$. At the centre there is a circle of radius $\xi_{1/2}$ about the origin of height $\alpha_0$ and pressure $\hat{P}_0$. In our simulations we typically used a discretization with $n=800$.

The annuli (and circle) heights change due to the flux at the edges, $\hat{Q}_{i\pm 1/2}$, giving evolution equations for the $\alpha_i$ in terms of the meniscus position $R_M$ and the neighbouring $\alpha_i$ and $\hat{P}_i$. The meniscus position evolves due to the (discretized) pressure gradient there.

The boundary conditions at $\xi=0$ are accounted for by the discretization at the centre, so we need only consider the boundary conditions at $\xi=1$. Applying these conditions fixes the value of $\alpha_n$, which is calculated by adding an additional `ghost' annulus outside the meniscus with height $\alpha_{n+1}$ and using central differences to discretize the two boundary conditions at each time step.

The solution to this series of ODEs in time, including the evolution of $R_M$, is found numerically from a given initial condition using MATLAB's inbuilt solver \texttt{ode15s}.

\section{Linear stability} \label{Ap:LinStab}

We consider small perturbations to a given equilibrium membrane shape, $\bar{H}(R)$, and corresponding meniscus radius, $\bar{R}$, at fixed values of the parameters $(\G, \Hinf)$. We take a standard ansatz of the form
\begin{equation}
\begin{aligned}
H(R,t) &= \bar{H}(R) + \varepsilon f(R) e^{\sigma t} \\
R_M(t) &= \bar{R} + \varepsilon e^{\sigma t}
\end{aligned}
\end{equation}
where $\varepsilon \ll 1$ and $\sigma$ is the growth rate of the perturbation.

The function $f(R)$ and the growth rate $\sigma$ are determined by linearizing (with respect to $\varepsilon$) the evolution equation \eqref{eq:PDE} and boundary conditions \eqref{eq:BCsymm}--\eqref{eq:BCmeniscus}, as well as volume conservation \eqref{eq:VolND}. The problem can be written in a more convenient form by introducing the function $I(R)=\int_0^R z f(z) \mathrm{d}z$, which measures the change in drop volume within $r<R$. We then obtain a pair of coupled ordinary differential equations for $f(R)$ and $I(R)$
\begin{align} 
\frac{1}{R} \frac{\mathrm{d}}{\mathrm{d}R}\left[ R \bar{H}^3 \frac{\mathrm{d}}{\mathrm{d}R} \left( \frac{1}{R}\frac{\mathrm{d}}{\mathrm{d}R} \left[ R \frac{\mathrm{d}f}{\mathrm{d}R} \right] \right) \right] + \sigma \G f = 0, \label{eq:LinStab1} \\
\frac{\mathrm{d}I}{\mathrm{d}R} = R f \label{eq:LinStab2}
\end{align} 
together with the six boundary conditions
 
\begin{align} \label{eq:LinStabBCs}
f'(0)=0, \qquad &f'''(0) =0, \qquad I(0)=0,  \nonumber \\
I(\bar{R}) &= -\bar{R} \bar{H}(\bar{R}), \nonumber \\
f'(\bar{R}) - &\frac{f(\bar{R})}{\bar{R} \log \bar{R}} = \frac{-2 \G}{\bar{H}(\bar{R})}, \\
f''(\bar{R}) + \frac{f'(\bar{R})}{\bar{R}} + &\frac{2 \G f(\bar{R})}{[\bar{H}(\bar{R})]^2} = \frac{-2 \G [\bar{H}(\bar{R})-\Hinf]}{[\bar{H}(\bar{R})]^2 \bar{R} \log \bar{R}}. \nonumber
\end{align}
These boundary conditions emerge from, respectively: symmetry at the origin, no radial flux at the origin, definition of $I$, volume conservation, fixed position at the clamp/gradient matching at the meniscus, and the curvature-imposed meniscus pressure.

The ODEs \eqref{eq:LinStab1} \& \eqref{eq:LinStab2} are solved numerically subject to the boundary conditions \eqref{eq:LinStabBCs} using MATLAB's inbuilt boundary value problem solver \texttt{bvp4c} to find the growth rate $\sigma$, given $(\G,\Hinf)$ and a valid equilibrium solution $\bar{H}(R)$. The dashed lines in fig.~\ref{fig:DynamicForce}a have slope corresponding to the growth rate determined in this way.

\section{Dimple drainage calculation}

\subsection{Similarity solution} \label{Ap:SimSoln}

We determine the pre-factors in the scaling relations \eqref{eq:DimpleScaling} by first finding a similarity solution for the membrane shape in the neck region (region II in fig.~\ref{fig:DimpleDiagram}). We note that the decrease in volume of the dimple controls the flux through this drainage region ${2 \pi C_\infty Q = -\dot{V}_{\text{dimple}}}$, and so from \eqref{eq:LocalFlux} the shape in this region must obey
\begin{equation}
		 H^3 \frac{\partial^3 H}{\partial X^3} = - \frac{\G}{4} C_\infty \dot{H}_0
\end{equation}
where we recall that $X=R-C_\infty$ is a local coordinate in the neck region.

Far from this region we match the solution to the curvature outside the dimple and the membrane slope immediately inside the dimple
\begin{equation}
\begin{aligned}
		\frac{\partial^2 H}{\partial X^2} &\to \G \Pi 	\qquad 	&\text{as } X \to +\infty \\
		\frac{\partial H}{\partial X} &\to -\frac{2 H_0}{C_\infty}		\qquad 	&\text{as } X \to -\infty.
\end{aligned}
\end{equation}

We look for a similarity solution in terms of a similarity variable
\begin{equation*}
	\eta = \frac{\G^{1/2} \Pi^{3/4}}{C_\infty^{1/2} }t^{1/4} (R-C_\infty),
\end{equation*} which is motivated by the scaling results \eqref{eq:DimpleScaling}. The similarity solution, $f(\eta)$, is defined by
\begin{equation*}
H = \frac{C_\infty}{ \Pi^{1/2}} ~t^{-1/2} f(\eta)
\end{equation*} and $H_0 = A ~ \G^{1/2} \Pi^{1/4} C_\infty^{3/2} ~t^{-1/4}$ where $A$ is a constant. Both $A$ and $f(\eta)$ are to be determined from the solution of the boundary value problem
\begin{equation}
\begin{aligned}
	f^3 f''' &= \frac{A}{16}, \\
	f' \to -2A \quad& \quad (\eta \to -\infty), \\
	f'' \to 1 \quad& \quad (\eta \to \infty).
\end{aligned}
\label{eq:simProb}
\end{equation}

Note that it is known \cite[][]{Jones1978} that there is a unique solution $g(z)$ to the problem $g^3g'''=1$ with boundary condition $g' \to -1$ as ${z \to -\infty}$ and $g'' \to k$ as ${z \to \infty}$ for some $k$. The constant $k$ is found numerically to be $k \approx 1.21$. We therefore seek to rescale our problem \eqref{eq:simProb} onto this classic problem by seeking a solution of the form $f(\eta) = \alpha g(\eta/\beta)$. We find that
\beq
\begin{aligned} \label{eq:SimSolnConsts}
&A = (2^9 k)^{-1/4}\approx0.20, 	\\
\alpha &= \frac{1}{2^7 A^2}, 	\qquad	\beta = \frac{1}{2^8 A^3}	
\end{aligned}
\eeq
and thus we have found a similarity solution for the profile at the dimple edge. See \cite{Jones1978} for a plot of the similarity solution $g(z)$.

\subsection{Adhesion force evolution} \label{Ap:Force Evolution}

To find the time-dependence of the adhesion force, we must first determine how the meniscus position and radius evolve. At leading order the membrane shape outside the dimple will behave quasi-statically, changing due to a small influx in volume as the fluid dimple drains. Similarly to the equilibrium contact solutions \eqref{eq:ContactCondition}--\eqref{eq:ContactVol}, we have three conditions relating the meniscus radius, height and the contact point but we replace \eqref{eq:ContactVol} by
\begin{equation}
1 - V_{\text{dimple}} = \frac{\pi \G}{H_M} \left[ \frac{R_M^4 - C^4}{4} + C^2 R_M^2 \log{\frac{C}{R_M}} \right]. \label{eq:DimpleContactVol}
\end{equation}

From the dimple scaling analysis \eqref{eq:DimpleVol}, we know that at a given time $t$ the volume of the dimple is approximately ${V_{\text{dimple}} \approx \frac{\pi}{2} A \G^{1/2} \Pi^{1/4} C_\infty^{7/2}  t^{-1/4}}$ where $A \approx 0.20$.

We expand the equations \eqref{eq:ContactCondition}, \eqref{eq:ContactGrad} \& \eqref{eq:DimpleContactVol} as follows: 
\begin{align*}
R_M &= R_M^\infty + \delta R_M, \\
H_M &= H_M^\infty + \delta H_M, \\
C &= C_\infty + \delta C
\end{align*}
with the leading terms (denoted with a super- or sub-script $\infty$) obeying the equilibria equations \eqref{eq:ContactCondition}--\eqref{eq:ContactVol} and the corrections (denoted by $\delta$) being small. Linearizing, we can write the resulting three linear equations for the first order corrections in the form $M\mathbf{x}=\mathbf{v}$ where the matrix $M$ is

\begin{widetext}
\begin{equation}
M =
\begin{pmatrix}
\G (R_M^\infty ~^2 - C_\infty ~^2) & -2 R_M^\infty H_M^\infty & - 2 \G R_M^\infty C_\infty \log (R_M^\infty / C_\infty) \\
\G  (R_M^\infty ~^2 - C_\infty ~^2 + 2 R_M^\infty ~^2 \log R_M^\infty) & R_M^\infty (\Hinf - 2 H_M^\infty) & - 2 \G R_M^\infty C_\infty \log R_M^\infty \\
2 \pi R_M^\infty H_M^\infty ~^2 & -1 & \pi \G C_\infty [R_M^\infty ~^2 - C_\infty ~^2 - 2 R_M^\infty ~^2 \log (R_M^\infty / C_\infty)]
\end{pmatrix}
\end{equation}
\end{widetext}
and the vectors $\mathbf{x}$ and $\mathbf{v}$ are
\begin{align}
\mathbf{x} =&
\begin{pmatrix}
\delta R_M \\ \delta H_M \\ \delta C
\end{pmatrix}, \\
\mathbf{v}& = -\frac{\pi}{2} A \G^{1/2} \Pi^{1/4}  C_\infty^{7/2} H_M^\infty ~t^{-1/4}
\begin{pmatrix}
0 \\ 0 \\ 1
\end{pmatrix}.
\end{align}

From this formulation, we see that knowing the equilibrium solutions, we can find the late-time leading order corrections to the radius and height of the meniscus, as well as the correction to the position of the contact point.

We then calculate the correction to the adhesion force from the expression
\[ F = \frac{2 \pi}{\G} \frac{H_M-\Hinf}{\log R_M}		\]
which, when linearized, gives the first order force correction
\begin{equation}
\delta F = \frac{2 \pi (\Hinf - H_M^\infty)}{\G R_M^\infty (\log R_M^\infty)^2}\delta R_M + \frac{2 \pi}{\G \log R_M^\infty} \delta H_M
\end{equation}
where $F_\infty$ is the equilibrium value of the force. Hence, we find that the leading order correction to the adhesion force is of size $t^{-1/4}$.

\section{Calculation of the dynamic work of separation} \label{Ap:Dynamic Work Calc}

To calculate the dynamic work of separation numerically, we use the numerical scheme described in Appendix \ref{Ap:Numerics}. These simulations begin with the membrane close to its equilibrium shape (the initial condition used for fig.~\ref{fig:DynamicEnergy} was found by evolving the system from the flat state $H=\Hinf$ for a dimensionless time $t=10^3$ with $\Hinf=5$, $\G=100$ both fixed). The value of the edge height $\Hinf$ is then increased at a given rate, $\dot{H}_\infty$.

The integral that defines the work of separation $\Wadh$ involves integrating to $\Hinf=\infty$; in practice, the dynamic numerics is terminated at a separation $H^*$, which is chosen to satisfy two requirements: (i)  the adhesive force at $H^*$ should be within $1\%$ of the corresponding equilibrium  value  and (ii)  the work of separation in moving from $H^*/5$ to $H^*$ is within 1\% of the corresponding quasi-static work of separation. For $\Hinf \geq H^*$ we therefore expect the adhesive force to remain very close to the quasi-static value and account for the work of separation in moving from  $H^*$ to $\infty$ as the value of the equilibrium work of separation from $H^\ast$ to $\infty$. For the results presented in fig.~\ref{fig:DynamicEnergy}b, a typical value was found to be ${H^*=100}$.


%

\end{document}